\documentclass[a4paper,12pt]{article}

\usepackage[top=25mm,bottom=25mm,left=25mm,right=25mm]{geometry}
\usepackage{amsmath}
\usepackage{amssymb}
\usepackage{mathtools}
\usepackage{cuted}
\usepackage{graphicx}
\usepackage{epstopdf}
\usepackage{url}
\usepackage{multicol}
\usepackage{cite}
\usepackage{url}
\usepackage{cleveref}
\usepackage{setspace}
\usepackage{floatrow}
\usepackage[rightcaption]{sidecap}
\usepackage{titlesec}
\titleformat{\section}{\bfseries\large\scshape\filcenter}{\thesection}{1em}{}
\titleformat{\subsection}{\bfseries\normalsize\scshape\filcenter}{\thesubsection}{1em}{}

\newcommand{\captionfonts}{\footnotesize}
\renewcommand\thesection{\Roman{section}}
\renewcommand\thesubsection{\Alph{subsection}}

\makeatletter
\long\def\@makecaption#1#2{
  \vskip\abovecaptionskip
  \sbox\@tempboxa{{\captionfonts #1: #2}}%
  \ifdim \wd\@tempboxa >\hsize
    {\captionfonts #1: #2\par}
  \else
    \hbox to\hsize{\hfil\box\@tempboxa\hfil}%
  \fi
  \vskip\belowcaptionskip}

\renewcommand\p@subsection{\thesection}
    
\makeatother

\begin{document}

\title{\textbf{\large{Predicting the Directional Transport of Multivalent Cargo from Position Dependent Binding and Unbinding Rates}}}

\author{\normalsize{L.S. Mosby$^{1, 2, 3}$, A. Straube$^1$, M. Polin$^{1, 2, 4}$} \\
	\small\textit{
		$^1$ Centre for Mechanochemical Cell Biology, University of Warwick, Coventry CV4 7AL, UK} \\
	\small\textit{
		$^2$ Physics Department, University of Warwick, Coventry CV4 7AL, UK} \\
	\small\textit{
		$^3$ Institute of Advanced Study, University of Warwick, Coventry CV4 7AL, UK} \\
		\small\textit{
		$^4$ Mediterranean Institute for Advanced Studies, IMEDEA, UIB-CSIC, Esporles, 07190, Spain}}

\date{}
\maketitle

\begin{abstract} 
	Multivalent cargo that can interact with substrates via multiple interaction sites exhibit shared characteristics despite being found in different systems at different length-scales. Here, a general analytical model has been developed that can describe the motion of multivalent cargo as a response to position dependence in the binding and unbinding rates of their interaction sites. Cargo exhibit both an effective diffusivity and velocity, which acts in the direction of increasing cargo-substrate binding rate and decreasing cargo-substrate unbinding rate. This model can reproduce  previously published experimental findings using only the binding and unbinding rate distributions of cargo interaction sites, and without any further parameter fitting. Extension of the cargo binding model to two dimensions reveals an effective velocity with the same properties as that derived for the $1$D case.
\end{abstract}

\vspace{1cm}

\section{Introduction} \label{Intro}

Multivalent cargo, defined as cargo that can interact with a substrate via multiple interaction sites, exist at a range of microscopic length-scales and exhibit a range of interesting dynamical phenomena. For example: small multivalent ligand molecules exhibit diffusive motion on receptor-functionalised surfaces \cite{Perl2011}; polymers consisting of many protein molecules interact and guide each other's growth via multiple crosslinkers \cite{Applewhite2010, Lopez2014, Forth2014, Alkemade2021}; chromosomes (collections of DNA and proteins) that can simultaneously interact with multiple polymers in cells are transported processively towards the cell poles during mitosis \cite{Gorbsky1987, Volkov2018}, and organelles tubulate in response to rapid interactions between their membrane-associated proteins and passing growing polymers \cite{WatermanStorer1995, WatermanStorer1998, Grigoriev2008}. Many of these examples describe the interactions between multivalent cargo and microtubules in cells, which are hollow cylindrical polymers that act as `tracks' for the directional transport of vesicles by motor proteins \cite{Nogales1999}. These cargo-microtubule interactions can occur directly, or can be mediated by highly specialised end-binding proteins (EBs) that transiently and preferentially bind to the unique structure at the growing ends of microtubules \cite{Su1995, Beinhauer1997, Renner1997, Tirnnauer1999, Nakagawa2000, Su2001, Bieling2007, Bieling2008}. Cargo motion driven by binding and unbinding events is distinct from that exhibited by monovalent particles that cannot move by this mechanism once bound. For example, EBs accumulate at growing microtubule ends only at the population-level due to their preferential binding dynamics \cite{Bieling2007, Bieling2008, Grigoriev2008, Honnappa2009, Applewhite2010, Gouveia2010, Jiang2012, Maurer2012, Roth2018}, whereas cargo that can interact simultaneously with multiple EBs are able to co-move with growing microtubule ends whilst maintaining at least one cargo-EB-microtubule linkage \cite{Rodriguez2020, Alkemade2021}.

Although the example systems listed above each consist of different interaction networks, they share the characteristics that each interaction site can interact only transiently with their corresponding substrate, and that relatively few interaction sites are required to generate cargo motion \cite{Perl2011, Zaytsev2013, Rodriguez2020, Alkemade2021}. It can also be predicted that the average dwell times of each type of cargo strongly depend on the number of their interaction sites that can bind to the substrate \cite{Klumpp2005, Erdmann2012, Zaytsev2013, Volkov2018}. It is therefore expected that the underlying mechanism by which the valency of cargo is coupled to their dynamical behaviour is the same for all of these systems, but no general model has yet been developed to show this.

Traditionally, the dynamics of cargo systems are modelled by studying the discrete binding dynamics of individual interaction sites \cite{Joglekar2002, Klumpp2005, Perl2011, Erdmann2012, Zaytsev2013, Rodriguez2020, Alkemade2021}, which often require numerical solvers due to their combinatorial complexity \cite{Errington2019}. Cargo dynamics can then be obtained directly using simulations \cite{Perl2011, Forth2014, Rodriguez2020}, or derived analytically from either the forces generated by the elongation of the (approximately `spring-like') linkers that connect the interaction sites to the main body of the cargo \cite{Joglekar2002, Erdmann2012}, or the free energy associated with the possible binding states available for the cargo \cite{Alkemade2021}. For these approaches, it is often difficult to extract analytical formulae that describe the coarse-grained motion of the cargo being studied. Alternatively, the forms of coarse-grained parameters describing cargo motion can be assumed, and parameter fitting can be used to calibrate their evolution based on experimental findings. Recent models have begun to investigate how position dependence in cargo-substrate interactions can result in the directional motion of cargo, but this is currently limited to step-like gradients in simplified interaction networks \cite{Alkemade2021}.

Here, a general model of cargo motion has been developed that only requires the binding and unbinding rate distributions of individual interaction sites as input parameters, such that predictions of coarse-grained dynamical parameters can be made without parameter fitting. Equations that describe cargo motion have been derived explicitly and show that cargo diffuse in the absence of position dependence in cargo-substrate binding and unbinding rates, but that they exhibit a deterministic effective velocity when position dependence in the rates is introduced. The analytical form of this effective velocity agrees with that obtained using stochastic cargo binding simulations, and can be used to show that cargo accumulate at regions of increased cargo-substrate binding rate. Cargo dynamics will be shown to fall into one of three regimes depending on their number of interaction sites: cargo with too few interaction sites exhibit dwell times too small to generate meaningful motion while bound; cargo with moderate numbers of interaction sites exhibit motion governed by both their effective velocities and diffusivities; and cargo with too many interaction sites exhibit approximately deterministic motion due to damping of their effective diffusivity. Together, results from modeling and simulations also suggest that the distribution of substrate-bound cargo interaction sites strongly affects the cargo's ability to diffuse on a substrate. In order to test this model in biologically-relevant conditions, experimental parameters from previously published works have been used to recapitulate the EB-mediated processive transport of cargo. Finally, the model has been expanded to two dimensions with the aim of describing more complex biological systems. The general model developed in this work is valid for any form of the binding and unbinding rate distribution, for arbitrarily many types of interaction.

\section{Results}

\subsection{The Cargo Binding Model} \label{sec:Analytics}

\begin{figure*}[b!]
\includegraphics[width=0.5\linewidth]{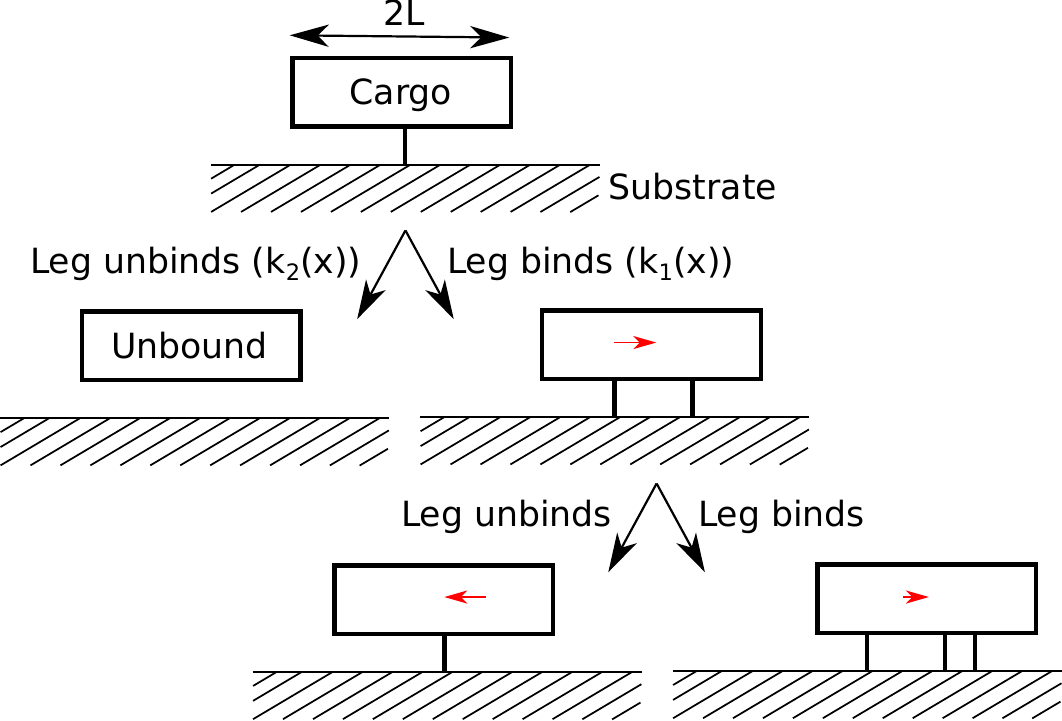}
	\caption{A cargo binding model where the centre position $x_a(t)$ of each $N$-legged cargo moves (red arrow) as a result of one of its legs binding to (leg added, $m=1$) or unbinding from (leg removed, $m=2$) a substrate with the position dependent interaction rate $k_m(x)$. The cargo is unbound when it no longer has any legs bound to the substrate.}
	\label{fig:Model}
\end{figure*}

It can be hypothesised that the bound motion of cargo that can interact with a substrate via multiple interaction sites (referred to as legs) originates from the ability of these legs to rapidly unbind and then rebind at a different position on the substrate (see Fig.(\ref{fig:Model})). In this case, each binding or unbinding event could result in the displacement of the centre position of the cargo $x_a(t)$, which has been defined as the averaged position of the cargo's legs that are currently bound to the substrate. This enforces the simple force equilibrium $\sum_{l=1}^{l=n} \kappa(x_l-x_a(t))=0$ for cargo with $n$ legs bound at positions in the set $\{ x_l \}(t)$ (where $\kappa$ is an effective spring constant). Cargo initially bind in the $n=1$ state and are unbound once $n=0$, such that their average dwell times can be calculated using previously published formulae \cite{Klumpp2005, Erdmann2012} (see Fig.(\ref{fig:TDwell}) in the \textit{Supplementary Information}).

In this model, each cargo leg can bind to ($m=1$) or unbind from ($m=2$) the substrate with the position dependent rates $k_m(x)$, but these rates need to be modified to take into account the size and shape of the cargo. This effect can be introduced through a binding rate distribution of the form $k_1(x|x_a(t),n)=k_1(x)S^N(x|x_a(t),n)$, where $S^N(x|x_a(t),n)$ is a normalised distribution dubbed the `shape factor'. For example, a shape factor of the form $S_0^N(x|x_a(t))=\theta(x+(x_a(t)-L))-\theta(x-(x_a(t)+L))$ (where $\theta(x)$ is the Heaviside step function) states that cargo legs can only bind to the microtubule at positions within the range $x_a(t)-L \leq x \leq x_a(t)+L$, defining a cargo width of $2L$. More complex shape factors can be used to investigate the effects of entropy limiting the extension of cargo legs and inhibiting their ability to bind far from $x_a(t)$. The model studied in this work assumes that timescales associated with the diffusion of cargo legs on the surface of cargo ($t_d\sim(2L)^2/D_l$) are much smaller than the binding timescales for these legs ($t_b\sim 1/\max(k_1(x))$), such that any leg can bind at any position on the substrate (limited by the shape factor $S^N(x|x_a(t),n)$) regardless of how long ago it may have unbound previously. Limits on the range of possible positions a cargo leg can bind to as a function of time since unbinding can be introduced through explicit time dependence in a shape factor for each leg.

Since cargo legs are stationary once bound, unbinding events can only occur from the positions in the set $\{ x_l \}(t)$. Although simple to implement computationally, this effect requires the introduction of the coarse-grained bound leg distribution $P_l^N(x|x_a(t),n)$ when trying to study these systems analytically. This normalised distribution describes the probability of one of the cargo's $n$ bound legs being at the position $x$ when the cargo centre position is $x_a(t)$, and modifies the unbinding rate distribution such that on average $k_2(x|x_a(t),n)=k_{off}(x)P_l^N(x|x_a(t),n)$. Importantly, if the shape distribution only permits binding of cargo legs within a specified range of positions, then the bound leg distribution will decay quickly at the extremities of this range.

In the presence of position dependent cargo leg binding and unbinding rates $k_{1,2}(x)$, cargo will exhibit position dependent average displacements and event rates. The $i^{th}$ moment of the cargo displacement distribution ($\lambda_m^{(i)}(x_a(t),n)$) and the average rate ($\bar{k}_m(x_a(t),n)$) associated with each type of event ($m=1,2$) can be calculated exactly via the equations,

\begin{equation}
\lambda_m^{(i)}(x_a(t), n) = \left( \frac{1}{n +\Delta_m} \right)^i \left( \frac{\int\limits_{b_l}^{b_u}\, \left([\Delta_m (x - x_a(t))]^i\, k_m(x|x_a(t),n) \right) dx}{\int\limits_{b_l}^{b_u}\, k_m(x|x_a(t),n) \, dx} \right),
\label{eq:DiscreteMoment}
\end{equation}

\begin{equation}
\bar{k}_m(x_a(t), n) = \left( \frac{N\,\delta_{m,1}-n\,\Delta_m}{b_u-b_l} \right) \int\limits_{b_l}^{b_u} \,k_m(x|x_a(t),n)\,dx,
\label{eq:DiscreteRate}
\end{equation}

\noindent where $\Delta_m=\delta_{m,1}-\delta_{m,2}$, $\delta_{i,j}$ is the Kronecker delta function, and $b_{u,l}$ are the upper and lower bounds of the averages defined by the shape factor and the bound leg distribution, which are set to $x_a(t)\pm L$ by substituting in the shape factor $S_0^N(x|x_a(t))$. The second term in eq.(\ref{eq:DiscreteMoment}) defines the mean $i^{\textnormal{th}}$ power of the difference between the positions of the binding or unbinding event and the cargo centre, and the scaling $\propto (1/(n+\Delta_m))^i$ enforces that the cargo centre position will exhibit smaller displacements when more legs are bound. In contrast, eq.(\ref{eq:DiscreteRate}) states that the rates of binding or unbinding events are averaged over the extent of the cargo, but that they increase proportionally with the number of legs available for each type of transition. Importantly, eq.(\ref{eq:DiscreteMoment} \& \ref{eq:DiscreteRate}) represent a Markovian system where the next binding or unbinding event of a cargo only depends on its current position and number of bound legs, despite this not being the case per cargo in stochastic binding simulations.

A third dynamical component ($m=3$) is introduced to take into account any deterministic cargo motion that is independent of position and $n$. This could be used to model cargo that bind to substrates that themselves can move \cite{Joglekar2002, Zaytsev2013, Volkov2018}, cargo that bind to motor proteins in cells (assuming that cargo velocity is independent of the number of associated motors) \cite{Klumpp2005}, or cargo motion in the rest frame of another deterministically moving object, such as a linearly growing microtubule end. This motion can be introduced analytically by assuming that all cargo legs are displaced by $\Delta x$ (resulting in $\lambda_3^{(i)}=(\Delta x)^i$) at a rate $\bar{k}_3$, but can alternatively be adapted to accept any displacement or wait-time distribution.

The PDF $P(x,t)$ describing the probability of finding cargo at the position $x$ at time $t$ can be derived for a population using the relation $P(x,t+dt)=\langle \delta(x-(x_a(t+dt))) \rangle$, assuming $dt$ is an infinitesimally small timestep (see \textit{Supplementary Information}) \cite{Williams2021}. From this, a Fokker-Planck equation can be derived of the form,

\begin{equation}
\frac{\partial P(x,t)}{\partial t} = \frac{\partial}{\partial x} \left[ D_{eff}(x) \frac{\partial P(x,t)}{\partial x} \right] - \frac{\partial}{\partial x} \left[ v_{eff}(x) P(x,t) \right] + k_{on}^{eff}(x) - k_{off}^{eff}(x),
\label{eq:FokkerPlanck}
\end{equation}

\noindent where $k_{on,off}^{eff}(x)$ are the position dependent effective binding and unbinding rates (respectively) of cargo modelled as single, composite bodies. The position dependent effective velocity ($v_{eff}(x)$) and diffusivity ($D_{eff}(x)$) in eq.(\ref{eq:FokkerPlanck}) are defined as,

\begin{equation}
v_{eff}(x) = \sum\limits_{n=1}^N \left[ P_n(x) \sum\limits_{m=1}^M \left( \bar{k}_m(x,n)\, \lambda_m^{(1)}(x,n) \right) \right] - \frac{\partial D_{eff}(x)}{\partial x} = S_\lambda^{(1)}(x) - \frac{\partial D_{eff}(x)}{\partial x},
\label{eq:Veff}
\end{equation}

\begin{equation}
D_{eff}(x) = \left( \frac{1}{2} \right) \sum\limits_{n=1}^N \left[ P_n(x) \sum\limits_{m=1}^M \left( \bar{k}_m(x,n)\, \lambda_m^{(2)}(x,n) \right) \right] = \frac{S_\lambda^{(2)}(x)}{2},
\label{eq:Deff}
\end{equation}

\noindent where $P_n(x)$ is the probability of cargo having $n$ legs bound at the position $x$ (see \textit{Supplementary Information}), and $M=3$ for a system with binding and unbinding events and a net velocity. The terms $S_\lambda^{(1,2)}(x)$ both consist of sums over their independent contributions from $M$ different types of events, averaged over the expected number of bound legs at each position. As a result, these terms can be rewritten as the products $S_\lambda^{(1,2)}(x)=k_t(x)\,\lambda^{(1,2)}(x)$, where $k_t(x)$ is the position dependent total event rate and $\lambda^{(1,2)}(x)$ are the position dependent mean and mean-squared displacements averaged over all types of events and numbers of legs bound (see \textit{Supplementary Information}). More complex cargo systems can be studied by introducing more types of event ($M>3$). A derivation of eq.(\ref{eq:FokkerPlanck}) for a system with periodic boundaries is included in the \textit{Supplementary Information}.

Substituting eq.(\ref{eq:DiscreteMoment}) into eq.(\ref{eq:Veff}) results in an effective velocity that acts in the direction of increasing binding rate ($k_1'(x|x_a(t),n)>0$, where $'$ signifies a derivative with respect to $x$) and decreasing unbinding rate ($k_2'(x|x_a(t),n)<0$). A statement about the motion of individual cargo cannot be made from eq.(\ref{eq:FokkerPlanck}-\ref{eq:Deff}) alone, since they include the continuum cargo dynamics that arise due to spatial gradients in $P(x,t)$. For this reason, a discrete-scale Langevin approach to the study of cargo motion is also required.

Cargo that obey eq.(\ref{eq:FokkerPlanck}-\ref{eq:Deff}) individually follow the Langevin equation (see \textit{Supplementary Information}),

\begin{equation}
dx(t) = S_\lambda^{(1)}(x) \, dt + \sqrt{S_\lambda^{(2)}(x)}\, dW(t),
\label{eq:Langevin}
\end{equation}

\noindent while bound, where $dx(t)=x(t+dt)-x(t)$ and $dW(t)$ is a Wiener process term that obeys $w(dt)=\int_{t}^{t+dt}dW(t')\sim \mathcal{N}(0,dt)$ and $\langle w(dt) \rangle = 0$ \cite{Risken1989, VanKampen1992, Paul2013}. The first term in eq.(\ref{eq:Langevin}) defines the deterministic motion of bound cargo due to gradients in the binding and unbinding rates of their legs, such that $S_\lambda^{(1)}(x)$ is the effective velocity exhibited by individual cargo while bound to a substrate. The second term in eq.(\ref{eq:Langevin}) instead describes cargo's stochastic motion. This means that the fixed points of cargo motion will occur at positions where $\langle dx(t) \rangle = S_\lambda^{(1)}(x) \, dt = 0$, such that,

\begin{equation}
\sum\limits_{n=1}^N \left[ P_n(x) \left( \bar{k}_1(x,n)\,\delta_1^{(1)}(x,n) + \bar{k}_2(x,n)\,\delta_2^{(1)}(x,n) \right) \right] + \bar{k}_3\,\Delta x = 0,
\label{eq:StableFixedPoint}
\end{equation}

\noindent and that these fixed points will be stable when $\partial \langle dx(t) \rangle/\partial x < 0$ (or $\partial S_\lambda^{(1)}(x)/\partial x < 0$). For example, $\bar{k}_3\,\Delta x = -v_{MT}$ for the case of EB-mediated cargo transport in the rest frame of a microtubule end growing with velocity $v_{MT}$, and in this case a fixed point arises when the effective velocity generated by the binding dynamics of cargo legs balances the `net velocity' due to microtubule growth.

\subsection{Probing Cargo Dynamics}

Although difficult to derive analytically, it is trivial to obtain the bound leg distribution $P_l^N(x|x_a(t),n)$ from stochastic cargo binding simulations. Cargo were simulated using the Gillespie algorithm \cite{Gillespie1976, Gillespie2007} on the periodic domain $x \in [-h,h]$ , and their dynamics were recorded and analysed (see \textit{Supplementary Methods}). The form of the bound leg distribution will vary depending on the shape factor of the cargo (assumed to be $S_0^N(x|x_a(t))$ for the following sections) and the underlying binding and unbinding rate distributions of the cargo's legs. Example bound leg distributions for use in eq.(\ref{eq:DiscreteMoment} \& \ref{eq:DiscreteRate}) are shown in Fig.(\ref{fig:Diffusivity}a \& \ref{fig:BoundLegDist}a), and have been fitted as described in the \textit{Supplementary Information}.

\begin{figure*}[b!]
\includegraphics[width=1.0\linewidth]{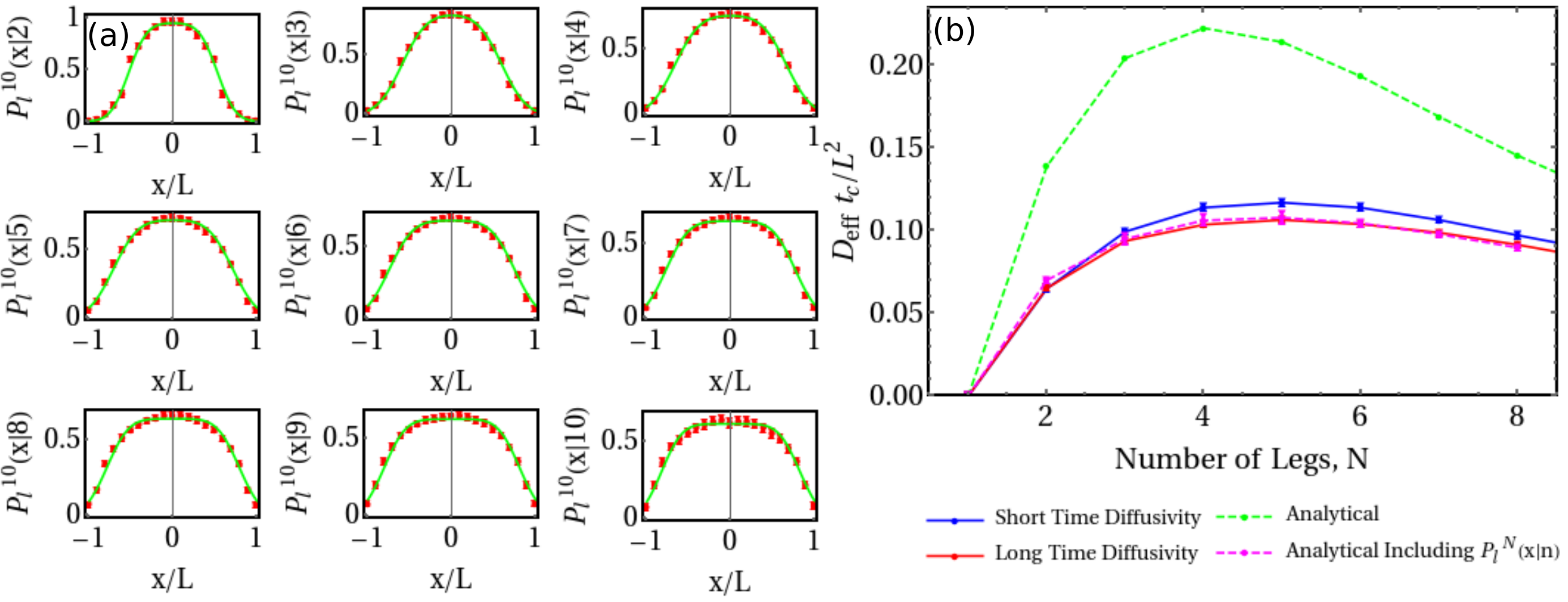}
	\caption{(a) Example bound leg distributions (red) for $10$-legged cargo averaged over cargo centre position (number of simulated cargo $n_{sim}=10\,000$). Fits (green) were calculated as described in the \textit{Supplementary Information}. (b) Cargo exhibit distinct short- and long-time diffusivities. Distributions from simulations were calculated from the gradients of mean-squared displacement distributions, as shown in Fig.(\ref{fig:MSD}) in the \textit{Supplementary Information} ($n_{sim}=100\,000$ for $1 \leq N \leq 7$, $n_{sim}=50\,000$ for $N=8$, $n_{sim}=25\,000$ for $N=9$, $n_{sim}=10\,000$ for $N=10$), and analytical distributions were obtained using eq.(\ref{eq:DiscreteMoment}, \ref{eq:DiscreteRate} \& \ref{eq:Deff}).}
	\label{fig:Diffusivity}
\end{figure*}

Although not being enforced explicitly, cargo simulations with $\bar{k}_3=0$ and position independent binding and unbinding rates reveal that cargo exhibit diffusive motion (see Fig.(\ref{fig:MSD}a) in the \textit{Supplementary Information}). Distinct short- and long-time diffusivities can be observed in Fig.(\ref{fig:Diffusivity}b) as a result of the binding dynamics of cargo, which could explain previously published experimental observations \cite{Alkemade2021}. Cargo initially bind in the $n=1$ state where the mean-squared displacement due to a binding event will be maximal (eq.(\ref{eq:DiscreteMoment}) shows $\lambda_1^{(2)}(x_a(t),n) \propto 1/(n+1)^2$), but cargo with $n=1,2$ will be more likely to unbind within short timescales than cargo with more legs bound. This means that for longer timescales the dynamics of cargo with more legs bound on average will dominate, as more of them will remain bound and contributing to the mean-squared displacement. The average diffusivity of bound cargo therefore decreases over time as the highly motile `small $n$' states become relatively less occupied (see Fig.(\ref{fig:AvN}) in the \textit{Supplementary Information}). The definition of short- and long-time are relative to a `separation timescale' $\lesssim t_c$ for $N > 2$ (see the crossover point in Fig.(\ref{fig:MSD}b) and the the plateau in Fig.(\ref{fig:AvN}) in the \textit{Supplementary Information}), which diverges as $N \rightarrow 2$ since $D_{short}(N=2) \equiv D_{long}(N=2)$ (see Fig.(\ref{fig:Diffusivity}b)). The diffusivities in Fig.(\ref{fig:Diffusivity}b) are non-monotonically increasing functions of $N$ due to competition between the increasing total rate of events occurring and the decreasing average displacement per event (see eq.(\ref{eq:DiscreteMoment} \& \ref{eq:DiscreteRate})).

It can be observed in Fig.(\ref{fig:Diffusivity}b) that the analytical treatment of eq.(\ref{eq:DiscreteMoment}, \ref{eq:DiscreteRate} \& \ref{eq:Deff}) generates effective diffusivities in agreement with the long-time diffusivities obtained from simulations. In this case the long-time average $P_n^c$ distribution has been used, which neglects the contributions of rebinding cargo (see \textit{Supplementary Information}). Unfortunately, this also neglects the short-time behaviour of cargo occupying highly motile `small $n$' states, so only the long-time diffusivity can be calculated in this way. Without the inclusion of the bound leg distribution in eq.(\ref{eq:DiscreteMoment} \& \ref{eq:DiscreteRate}) the analytically-derived effective diffusivity is incorrect by approximately a factor of two (see Fig.(\ref{fig:Diffusivity}b)).

\begin{figure*}[b!]
\includegraphics[width=1.0\linewidth]{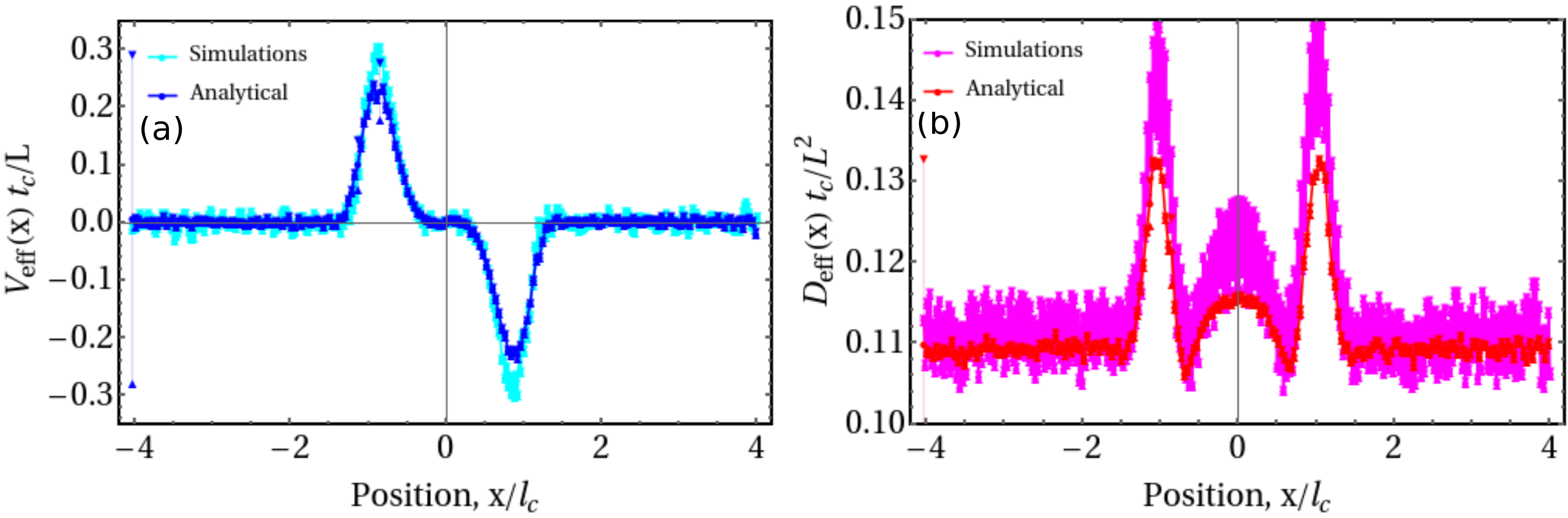}
	\caption{(a) Effective velocities, and (b) effective diffusivities, derived by substituting the results of simulations (blue, red, $n_{sim}=250\,000$) or eq.(\ref{eq:DiscreteMoment}\&\ref{eq:DiscreteRate}) (cyan, magenta) into eq.(\ref{eq:Veff}) or eq.(\ref{eq:Deff}) respectively. Both effective velocity distributions show that cargo are `attracted' towards the region of increased cargo-substrate binding rate.}
	\label{fig:Velocity}
\end{figure*}

It can be predicted using eq.(\ref{eq:DiscreteMoment}, \ref{eq:DiscreteRate} \& \ref{eq:Veff}) that $v_{eff}(x)=\bar{k}_3\,\Delta x$ for systems with position independent binding and unbinding rates. Now, consider the binding rate distribution,

\begin{equation}
k_1(x)\,t_c = \left( \frac{4}{1+\exp\left( (x/2)^6 \right)} \right) + 1,
\label{eq:PosDepRates}
\end{equation}

\noindent where the characteristic timescale $t_c=1/k_2$ is a constant, and the characteristic lengthscale $l_c=2(\ln(39))^{1/6} = 2.48...$ is the distance from the origin to the position where
$k_1(x) < 1.1\,k_2$ for the first time. The distribution in eq.(\ref{eq:PosDepRates}) has a region of increased binding rate at the centre of the domain and a region of constant binding rate outside this region (see Fig.(\ref{fig:BindingRate}) in the \textit{Supplementary Information}). It can be observed in Fig.(\ref{fig:Velocity}a) that eq.(\ref{eq:PosDepRates}) results in simulated cargo exhibiting an effective velocity that `attracts' them towards the central region. This can be true even when $\bar{k}_3\,\Delta x \neq 0$ (see Fig.(\ref{fig:NonZeroVMT}a) in the \textit{Supplementary Information}), in which case the stable fixed points of bound cargo motion can be predicted using eq.(\ref{eq:StableFixedPoint}). Also plotted in Fig.(\ref{fig:Velocity}a) is the analytical effective velocity distribution, calculated using eq.(\ref{eq:DiscreteMoment}, \ref{eq:DiscreteRate} \& \ref{eq:Veff}) and the bound leg distributions shown in Fig.(\ref{fig:BoundLegDist}a) in the \textit{Supplementary Information}, which is very similar to that obtained using simulations. Similar agreement can be observed between the effective diffusivities calculated using each method (see Fig.(\ref{fig:Velocity}b)). Discrepancies between the distributions calculated analytically and those obtained using simulations are again the result of eq.(\ref{eq:DiscreteMoment}, \ref{eq:DiscreteRate} \& \ref{eq:Veff}) calculating only the long-time behaviour of cargo when averaged over $P_n(x)$.

\begin{figure*}[t!]
\includegraphics[width=\linewidth]{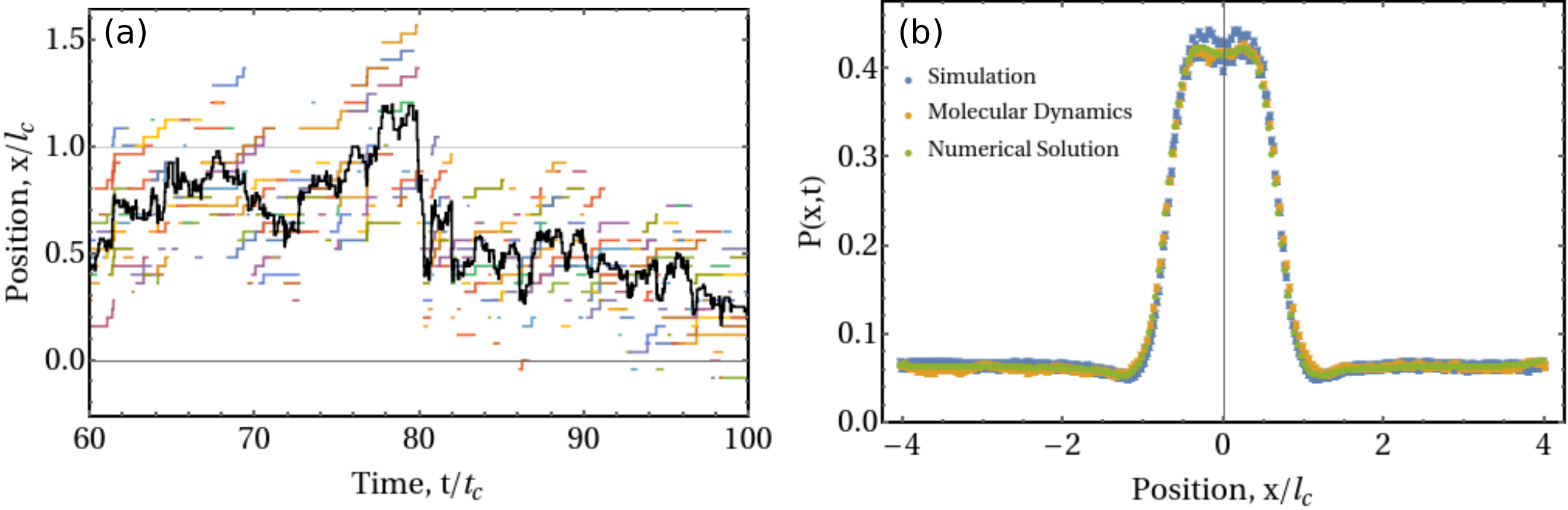}
	\caption{(a) Section of an individual cargo track (black) showing the positions of each leg while bound (coloured) when subject to $\bar{k}_3\,\Delta x=0.1\,L/t_c$. Cargo legs that unbind near $x=l_c$ preferentially rebind within the central region of increased cargo-substrate binding rate defined by eq.(\ref{eq:PosDepRates}) (see displacements near $t/t_c \simeq 80$). (b) The PDF $P(x,t)$ describing the probability of finding cargo at the position $x$ averaged over all simulation time $t$ exhibits a peak in the central region. Stochastic simulation results ($n_{sim}=250\,000$) agree with those obtained using molecular dynamics simulations ($dt=0.1$, $t_{max}=1\,000\,000$) and by numerically solving eq.(\ref{eq:FokkerPlanck}) using an analytically-derived effective velocity distribution. Data comprising the distribution obtained from stochastic simulations has been averaged over adjacent sites, but still exhibits fluctuations on a scale $\Delta x/(2l_c)$ (where $\Delta x$ is the separation between grid sites where cargo legs can bind) since cargo with $n=1$ can only occupy sites at integer multiples of $\Delta x/l_c$.}
	\label{fig:PDF}
\end{figure*}

Studying the motion of individual simulated cargo has revealed that their effective velocity is the result of the preferential binding of cargo legs within the central region of increased cargo-substrate binding rate (see Fig.(\ref{fig:PDF}a)). Despite the importance of including the bound leg distribution when calculating the effective diffusivity of cargo (see Fig.(\ref{fig:Diffusivity}b)), the component of the effective velocity due to unbinding events is negligible compared to the component due to binding events (see Fig.(\ref{fig:VelocityComponents} \& \ref{fig:NonZeroVMT}c) in the \textit{Supplementary Information}). The effect of the effective velocity is emphasised by the increased average dwell time of cargo within the central region, and together these effects result in a peak in $P(x,t)$ averaged over time (see Fig.(\ref{fig:PDF}b)). Using the $v_{eff}(x)$ and $D_{eff}(x)$ distributions derived analytically (see Fig.(\ref{fig:Velocity})), molecular dynamics simulations have been carried out using the Langevin dynamics defined in eq.(\ref{eq:Langevin}). These simulations generate a distribution $P(x,t)$ in agreement with those obtained using stochastic cargo binding simulations, and by numerically solving eq.(\ref{eq:FokkerPlanck}) using the same analytically-derived effective velocity distribution, as shown in Fig.(\ref{fig:PDF}b). In the case that $\bar{k}_3\,\Delta x \neq 0$, the distribution $P(x,t)$ is skewed in the direction of positive $\Delta x$ (see Fig.(\ref{fig:NonZeroVMT}b) in the \textit{Supplementary Information}).

\subsection{Behaviour of the Effective Velocity}

The explicit dependence of $S_\lambda^{(1)}(x)$ on the underlying rate distributions $k_{1,2}(x)$ can be approximated for cargo much smaller than the characteristic length scales associated with variation in these distributions (such that $L \ll (k_{1,2}(x)-k_{1,2}(x_a(t)))/k_{1,2}'(x_a(t))$ for the shape factor $S_0^N(x|x_a(t))$). In this case, the rates can be approximated to first order in $x$ as $k_{1,2}(x) \simeq k_{1,2}(x_a(t)) + (x-x_a(t))\,k_{1,2}'(x_a(t))$. Using these assumptions, the effective velocity defined in eq.(\ref{eq:Veff}) can be rewritten as,

\begin{equation}
\begin{aligned}
S_\lambda^{(1)}(x) & \simeq \sum\limits_{n=1}^N \left[ P_n(x) \left( \left( \frac{(N-n)L^2}{3(n+1)} \right) k_1'(x) - \left( \frac{n}{2L(n-1)} \right) k_2(x) \, I(y|x,n) \right) \right] \\
& \qquad + \bar{k}_3\,\Delta x,
\end{aligned}
\label{eq:ApproximateVeff}
\end{equation}

\noindent where $I(y|x,n)=\int_{x-L}^{x+L} ((y-x) P_l^N(y|x,n))dy$ is the average difference between the unbinding position of a cargo leg and the cargo centre position due to only variation in the bound leg distribution. The first term in eq.(\ref{eq:ApproximateVeff}) predicts that the component of the effective velocity due to binding events is proportional to the local gradient in the binding rate distribution, whereas the second term predicts that the component due to unbinding events is dominated by variation in the bound leg distribution.

It can be shown that the component of eq.(\ref{eq:ApproximateVeff}) due to binding events dominates over those resulting from unbinding events and gradients in the effective diffusivity for the cargo-substrate interaction rates defined in eq.(\ref{eq:PosDepRates}) (see Fig.(\ref{fig:VelocityComponents} \& \ref{fig:NonZeroVMT}c) in the \textit{Supplementary Information}). In this case, eq.(\ref{eq:ApproximateVeff}) predicts that $S_\lambda^{(1)}(x) \propto L^2\,k_1'(x)$, and that $S_\lambda^{(1)}(x)$ increases monotonically as a function of $N$ until it plateaus. In the rest frame of another deterministically moving object ($\bar{k}_3\,\Delta x \neq 0$), this means that cargo have the greatest likelihood of co-moving with this object at the position where $k_1'(x)$ is maximal, not where $k_1(x)$ peaks. In the case of cargo permanently associated to multiple EBs, this predicts that cargo would lag behind the peak of the EB comet distribution. However, the position dependence of $k_{on,off}^{eff}(x)$ in eq.(\ref{eq:FokkerPlanck}) would also influence the position of the observed intensity maximum for cargo in experiments when not at the single-molecule scale.

In the case of larger cargo with $2L \gtrsim l_c$ the assumptions used above are not valid, and it can be instead be assumed that the integral $\int_{x-L}^{x+L}\, ((y - x) k_1(y|x,n)) dx$ in eq.(\ref{eq:DiscreteMoment}) no longer varies significantly as a function of $L$ as variations in the binding rate distribution are averaged out over the extent of the cargo. This results in an effective velocity that approximately varies as $v_{eff}(x) \propto 1/L$. It is possible for large cargo to overhang the edges of non-periodic substrates, in which case they will `observe' an infinitely steep change in the binding and unbinding rates of their legs ($k_{1,2}(x>x_{edge})=0$). Assuming there is no other local variation in $k_{1,2}(x<x_{edge})=k_{1,2}$, the formalism in eq.(\ref{eq:DiscreteMoment}-\ref{eq:Deff}) can be used to show that cargo overhanging the edges of a substrate exhibit an effective velocity,

\begin{equation}
\begin{aligned}
    S_\lambda^{(1)}(x) & = \sum\limits_{n=1}^{N} \left[ P_n(x) \left( \frac{k_1(N-n)}{4L(n+1)} \right) \left( (x_{edge} - x)^2 - L^2 \right) - \left( \frac{k_2\,n}{2L(n-1)} \right) J(y|x,n) \right] \\
    & \qquad + \bar{k}_3\,\Delta x,
    \label{eq:EdgeVeff}
\end{aligned}
\end{equation}

\noindent where $J(y|x,n)=\int_{x-L}^{x_{edge}} ((y-x) P_l^N(y|x,n))dy$. The component of the effective velocity in eq.(\ref{eq:EdgeVeff}) resulting from binding events always acts towards the substrate (in the negative $x$-direction in this case), whereas the component resulting from unbinding events always acts away from the substrate (in the positive $x$-direction in this case) due to asymmetry in the bound leg distribution. Since the effective velocity due to binding events dominates over that resulting from unbinding events (see Fig.(\ref{fig:VelocityComponents} \& \ref{fig:NonZeroVMT}c) in the \textit{Supplementary Information}), cargo exhibit an effective velocity that tries to maximise their overlap with the substrate. According to eq.(\ref{eq:StableFixedPoint}), the effective velocity in eq.(\ref{eq:EdgeVeff}) could result in cargo exhibiting a stable fixed point in their motion near to the edge of the substrate when $\bar{k}_3\,\Delta x \neq 0$. This effect has been observed experimentally for actin filaments that interact with microtubules \cite{Alkemade2021}, but can also be used to predict that multivalent cargo can track the shrinking ends of depolymerising microtubules \cite{Volkov2018}.

\subsection{Using Experimentally-Derived Input Parameters}

\begin{figure*}[t!]
\includegraphics[width=\linewidth]{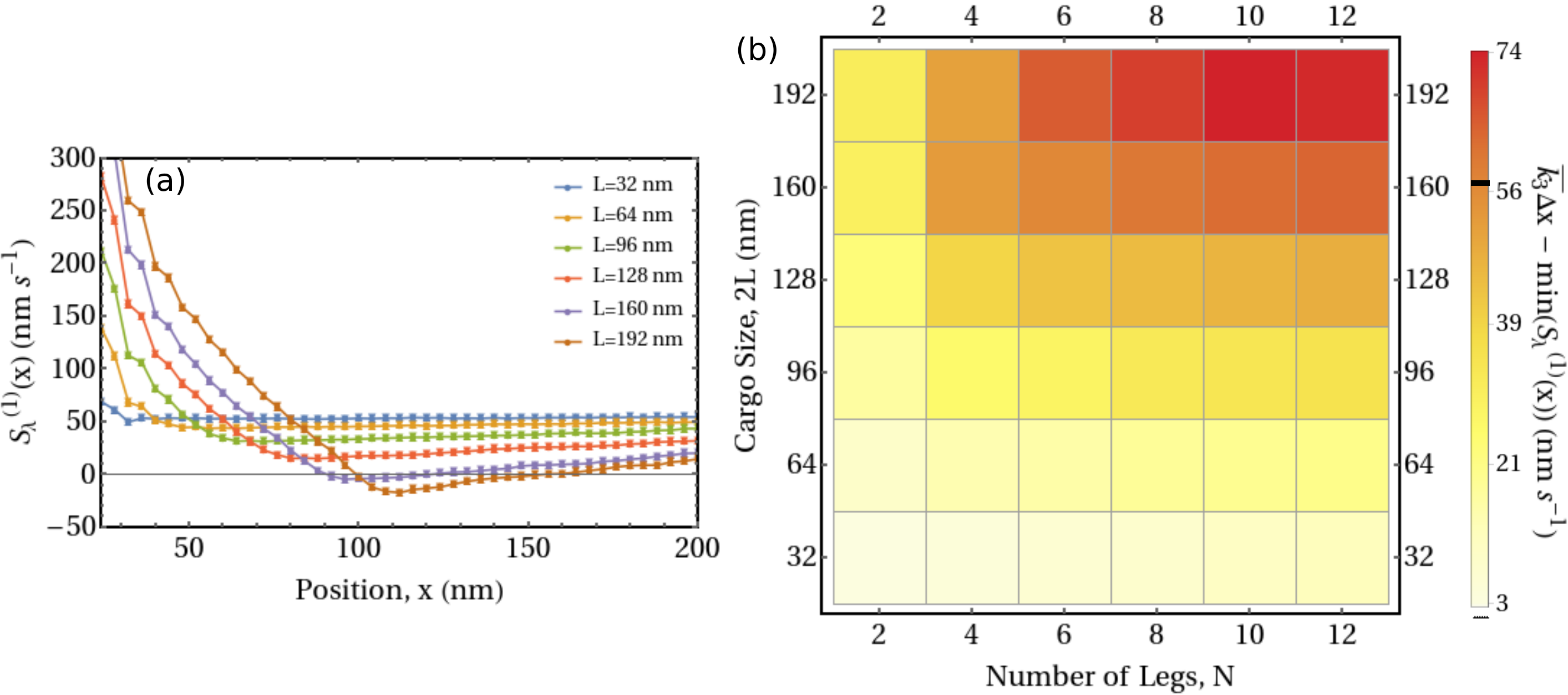}
	\caption{(a) The position dependent effective velocity exhibited by bound cargo permanently bound to $N=10$ EBs obtained from stochastic cargo binding simulations ($n_{sim}=50\,000$) using input parameters derived from previously published experimental data (see tables \ref{tab:ExperimentalModelValues} \& \ref{tab:ModifiedModelValues} in the \textit{Supplementary Information}). The microtubule edge is set at $x_{edge}=0$ nm, and stable fixed points are exhibited by $10$-legged cargo of sizes $2L \geq 160$ nm. (b) A heatmap showing the maximum effective velocity generated by binding or unbinding events in the direction of the growing microtubule end ($n_{sim}=100\,000$ for $2 \leq N \leq 8$, $n_{sim}=50\,000$ for $N = 10$, and $n_{sim}=25\,000$ for $N = 12$). Cargo can co-move with growing microtubule ends when $\bar{k}_3\,\Delta x - \min(S_\lambda^{(1)}(x))>57$ nm s${}^{-1}$ (black line on colour bar, corresponding to the microtubule growth velocity (see table \ref{tab:ExperimentalModelValues})).}
	\label{fig:ExpSimulations}
\end{figure*}

Experiments have shown that beads coated in EB binding domains can track the growing ends of microtubules in the presence of EBs \cite{Rodriguez2020, Alkemade2021}, but a model describing these dynamics has not yet been developed. In order to test whether the model presented in this work can reproduce the dynamics of multivalent cargo in biologically-relevant conditions, parameters describing EB-microtubule interactions obtained from experiments have been converted into input binding and unbinding rate distributions for stochastic cargo binding simulations \cite{Lodish2000, Gouveia2010, Buey2011, Maurer2014, Zhang2018, Roostalu2020, Song2020, Rodriguez2020} (see \textit{Supplementary Information}). In this section, simulations of cargo permanently bound to multiple EBs have been carried out, with results shown in Fig.(\ref{fig:ExpSimulations}). The variable $q$ has been used to sweep through possible values of the EB-microtubule interaction volume (see \textit{Supplementary Information}), such that the average number of bound legs for cargo increases monotonically as a function of $q$.

It can be observed in Fig.(\ref{fig:ExpSimulations}) that the model presented in this work predicts the tracking of growing microtubule ends by cargo in biologically-relevant conditions (see table \ref{tab:ExperimentalModelValues} in the \textit{Supplementary Information}). Using eq.(\ref{eq:StableFixedPoint}), stable fixed points were observed in the motion of $(N \geq 6)$-legged cargo of sizes $2L \geq 192$ nm and of $(N \geq 8)$-legged cargo of sizes $2L \geq 160$ nm. The effective velocity $S_\lambda^{(1)}(x)$ increases monotonically as a function of $N$ until it plateaus, and approximately quadratically as a function of $L$ for comparatively small cargo, as predicted by eq.(\ref{eq:ApproximateVeff}) (see Fig.(\ref{fig:ExpBreakdown}) in the \textit{Supplementary Information}). The sizes of cargo shown to track growing microtubule ends in Fig.(\ref{fig:ExpSimulations}) are of the same order of magnitude as the size of beads previously shown to exhibit these dynamics in experiments \cite{Rodriguez2020, Alkemade2021}, which may exhibit practical diameters of up to $\sim 60-70$ nm after considering the sizes of EB-cargo linkers ($\sim 10$ nm) and EBs ($\rho_{EB}^{max} = 13.6$ nm \cite{Buey2011}). Unlike for previously published experimental data, where it was not possible to predict the number of cargo-associated EBs that could interact with microtubules, Fig.(\ref{fig:ExpSimulations}) suggests that a surprisingly small number of permanently associated EBs are required to stimulate cargo transport. The steep velocity profile observed in Fig.(\ref{fig:ExpSimulations}) at positions close to the edge of the microtubule ($x=0$ nm) is generated by the same mechanism as the effective velocity in eq.(\ref{eq:EdgeVeff}).

\subsection{Extending the Cargo Binding Model to Two Dimensions}

In order to increase its generality, the cargo binding model defined by eq.(\ref{eq:DiscreteMoment}-\ref{eq:Deff}) has been expanded to $2$D to describe more complex cargo-substrate interaction networks. In this case, the PDF $P(\underline{x},t)$ describing the probability of finding cargo at the position $\underline{x}$ at time $t$ can be derived using the relation $P(\underline{x},t+dt)=\langle \delta(\underline{x} - \underline{x}_a(t+dt)) \rangle$ for a $2$D cargo centre position $\underline{x}_a(t)$. The resulting Fokker-Planck equation describing $2$D cargo motion can be written (see \textit{Supplementary Information}),

\begin{equation}
\begin{aligned}
\frac{\partial P(\underline{x},t)}{\partial t} & = \left( \frac{1}{2} \right) \underline{\nabla}\cdot \left[ \begin{pmatrix}
S_\lambda^{(2,x)}(\underline{x})\, \partial_x + S_\lambda^{(1,xy)}(\underline{x})\, \partial_y \\
S_\lambda^{(2,y)}(\underline{x})\, \partial_y + S_\lambda^{(1,xy)}(\underline{x})\, \partial_x
\end{pmatrix} P(\underline{x},t) \right] \\
& \qquad - \underline{\nabla}\cdot\left( \underline{v_{eff}}(\underline{x}) P(\underline{x},t) \right) + k_{on}^{eff}(\underline{x}) - k_{off}^{eff}(\underline{x}),
\end{aligned}
\label{eq:2DFokkerPlanck}
\end{equation}

\begin{equation}
\underline{v_{eff}}(\underline{x}) = \begin{pmatrix}
S_\lambda^{(1,x)}(\underline{x}) - \left( \frac{1}{2} \right)\left( \partial_x S_\lambda^{(2,x)}(\underline{x}) + \partial_y S_\lambda^{(1,xy)}(\underline{x}) \right) \\
S_\lambda^{(1,y)}(\underline{x}) - \left( \frac{1}{2} \right)\left( \partial_x S_\lambda^{(1,xy)}(\underline{x}) + \partial_y S_\lambda^{(2,y)}(\underline{x}) \right)
\end{pmatrix},
\label{eq:2DVeff}
\end{equation}

\noindent where $\partial_z \equiv \partial/\partial z$, and $S_\lambda^{(i,z)}(\underline{x})$ are as defined in eq.(\ref{eq:Veff} \& \ref{eq:Deff}) with moments of the cargo displacement distribution $\lambda_m^{(i,z)}(\underline{x}, n)$ averaged over the $2$D position variables $z^i$. The first term in eq.(\ref{eq:2DFokkerPlanck}) describes the effective diffusivity of cargo in $2$D, and so is calculated from the second moments of the cargo displacement distribution. The effective velocity defined in eq.(\ref{eq:2DVeff}) is instead calculated from the first moments of the cargo displacement distribution and terms related to gradients of the effective diffusivity, as was the case in eq.(\ref{eq:Veff}) for the $1$D cargo model.

For a $2$D system, cargo legs exhibit the baseline interaction rates $k_{m}(\underline{x})$, which must be corrected using a $2$D shape factor $S^N(\underline{x}|\underline{x}_a(t),n)$ and bound leg distribution $P_l^N(\underline{x}|\underline{x}_a(t),n)$. Once again, the simple shape factor $S_0^N(\underline{x}|\underline{x}_a(t))$ can be introduced that limits the possible binding positions of cargo legs to within the range $\sqrt{(x-x_a(t))^2+(y-y_a(t))} \leq L$. Using these definitions, equations for the moments of the cargo displacement distribution and the average rate associated with each type of event, equivalent to eq.(\ref{eq:DiscreteMoment} \& \ref{eq:DiscreteRate}) for the $1$D cargo model, can be derived (see the \textit{Supplementary Information}). Together, eq.(\ref{eq:2DFokkerPlanck}, \ref{eq:2DVeff} \& \ref{eq:DiscreteMomentX}-\ref{eq:2DDiscreteRate}) can be used to fully describe a $2$D cargo binding system. For example, it can once again be derived that cargo exhibit an effective velocity in the direction of increasing cargo-substrate binding rate and decreasing cargo-substrate unbinding rate, even for $2$D systems.

\section{Discussion}

Although the mechanisms driving many interesting dynamical systems can be explained via the interactions of multivalent cargo, a general model has not yet been developed that can decribe the plethora of phenomena observed for such systems. This model would need to be able to describe both the diffusive motion of cargo that exhibit position independent cargo-substrate interaction rates \cite{Perl2011}, and the processive motion of cargo that either interact with moving substrates or exhibit position dependent cargo-substrate interaction rates \cite{Joglekar2002, Zaytsev2013, Volkov2018, Rodriguez2020, Alkemade2021}, while also being extendable to higher valency cargo and systems with more complex interaction networks. In this work, equations that describe cargo motion have been derived explicitly from basic principles involving binding and unbinding events, and it has been shown that gradients in cargo-substrate interaction rates govern the direction and magnitude of the effective velocity exhibited by cargo. With only binding and unbinding rate distributions derived from experimental parameters as inputs, this cargo binding model has been shown to be able to qualitatively reproduce experimentally observed phenomena without any further parameter fitting.

It has recently been reported that the mean transport time of actin filaments (corresponding to the average amount of time they co-move with the growing microtubule end) exhibits a maximum as a function of filament length \cite{Alkemade2021}. This agrees with the results of the cargo model presented in this work, where it is expected that competition between the $v_{eff}(x) \propto L^2$ (see eq.(\ref{eq:ApproximateVeff})) and $v_{eff}(x) \propto 1/L$ regimes will limit the size of cargo observed inside cells. However, for large cargo it is possible that the assumption that the timescales associated with the diffusion of cargo legs on the surface of cargo are much smaller than the binding timescales for these legs ($t_d \ll t_b$) breaks down. In this case, introducing explicit time dependence in a shape factor for each leg is expected to further inhibit the diffusive and processive motion of large cargo, such that $v_{eff}(x) \propto L^{-\alpha}$ with $\alpha>1$. It can also be observed in Fig.(\ref{fig:ExpSimulations}a) that the predicted distance between the edge of a microtubule and the position of the stable fixed point of cargo motion increases with increasing cargo size. This is expected to be the result of competition between the $v_{eff}(x) \propto k_1'(x)$ (see eq.(\ref{eq:ApproximateVeff})) and $v_{eff}(x) \propto -((x_{edge}-x)^2 - L^2)$ (see eq.(\ref{eq:EdgeVeff}), with opposite sign due to the orientation of the microtubule in Fig.(\ref{fig:ExpSimulations})) regimes near the microtubule edge. Further experimental investigation is required to elucidate the size dependence of cargo motion.

Using the estimation method presented in the work by Alkemade \textit{et al.} \cite{Alkemade2021}, the effective forces acting on the cargo that generate the effective velocity and diffusivity distributions in Fig.(\ref{fig:ExpSimulations}a \& \ref{fig:ExpDiffusivities}a) can be calculated as $F(x) \simeq (k_B\,T\,v_{eff}(x))/D_{eff}(x)$. At the stable fixed points of cargo motion in Fig.(\ref{fig:ExpSimulations}a), this method predicts forces of $\sim 0.1$ pN (assuming a temperature of $37$ ${}^o$C), which is of the same order of magnitude as those obtained from optical tweezer experiments \cite{Rodriguez2020, Alkemade2021}, and approximately an order of magnitude smaller than those generated by individual motor proteins \cite{Meyhofer1995}. Assuming that motor proteins travel at velocities comparable to the microtubule growth velocity (the velocity exhibited by cargo while co-moving with the growing microtubule end) \cite{Lipka2016, Rickman2017, Zwetsloot2018, Roostalu2020}, and that the effective drag coefficients of cargo scale $\propto r_{cargo}$ (in accordance with Stokes' drag force), it can be predicted that the maximum size of cargo that can be transported by EB-mediated mechnanisms is $r_{tip}^{max} = (F_{tip}^{max}/F_{motor}^{max})r_{motor}^{max} \sim 0.1\,r_{motor}^{max}$. Of interest for future study is whether the binding and unbinding rates defined in tables \ref{tab:ExperimentalModelValues} \& \ref{tab:ModifiedModelValues} are affected by the association of EBs to cargo. For example, it can be predicted that association with microtubule-bound cargo may reduce the average EB-microtubule distance, and that this could increase the average amount of time an EB spends within the interaction volume of the substrate. Similarly, steric effects could reduce the freedom of cargo-EB linkers and inhibit the interactions between cargo-bound EBs and the microtubule.

Although the $1$D cargo binding model developed in this work has been shown to predict the tracking of growing microtubule ends by cargo permanently coated in EBs, the extension of the model to two dimensions allows for the implementation of a $2$D shape factor that takes into account the curvature of the cylindrical microtubule substrate. Alternatively, this $2$D model could provide a novel route for studying cell locomotion. It should be noted that eq.(\ref{eq:FokkerPlanck}) with $k_{on,off}^{eff}(x)=0$ has already been used to study the motion of passive particles in active-passive systems \cite{Williams2021}.

In conclusion, a general model for the motion of multivalent cargo bound to substrates has been derived that can both qualitatively and quantitatively reproduce experimentally observed phenomena. It has been shown that cargo exhibit an effective velocity that acts in the direction of increasing cargo-substrate binding rate and decreasing cargo-substrate unbinding rate, and that the magnitude of this effective velocity is approximately proportional to the local gradient of the binding rate distribution for comparatively small cargo. This work builds upon the results of previously published models by simultaneously deriving discrete and continuum-level analytics that can predict experimentally observable phenomena, and that require only the binding dynamics of individual EBs as inputs. The general model derived in this work has potential applications for many areas of biophysics research where protein or cell motion is the result of complex networks of binding dynamics.

\section*{Materials and Methods}

\subsection{Simulation Methods}

Stochastic cargo binding simulations were implemented in MATLAB using the Gillespie algorithm \cite{Gillespie1976, Gillespie2007} to probe the system state in continuous time (see \textit{Supplementary Methods}). Molecular dynamics simulations were also implemented in MATLAB, but instead updated cargo positions according to the Langevin equation defined in eq.(\ref{eq:Langevin}) using a forwards Euler scheme. Wiener process displacements were calculated using inverse transform sampling and binding dynamics were introduced by randomising the position of the cargo within the periodic domain with rate $k_{ran}(x) = 1 / t_{dwell}(x)$ (calculated by substituting the local binding and unbinding rates into a previously published average dwell time formula \cite{Klumpp2005, Erdmann2012}).

\subsection{Numerical Methods}

Numerical solutions to the Fokker-Planck equation defined in eq.(\ref{eq:FokkerPlanck}) were obtained using the built-in MATLAB function \url{pdepe()}.

\section*{Acknowledgements}

We acknowledge support from Leverhulme Trust Research Project Grant RPG-2016-260 (AS, MP, LSM), the Wellcome Trust Investigator Award 200870/Z/16/Z (AS) and the Ramo?n y Cajal Program (RYC-2018-02534; MP). MP and AS designed the project; LSM, MP and AS developed the mathematical model; LSM wrote the simulation code and manuscript. We would like to thank the Scientific Computing RTP at the University of Warwick for maintaining the HPC systems used to run simulations.

\setcounter{figure}{0}
\setcounter{subsection}{0}
\setcounter{equation}{0}
\setcounter{table}{0}
\renewcommand{\thefigure}{S\arabic{figure}}
\renewcommand{\thetable}{S\arabic{table}}
\renewcommand{\theequation}{S\arabic{equation}}
\renewcommand{\thesection}{} 
\renewcommand{\thesubsection}{\arabic{subsection}} 

\newpage

\section*{Supplementary Information}

\subsection{Deriving the Fokker-Planck Equation Describing Cargo\\ Motion in $1$D}

The following derivation of the Fokker-Planck equation to describe continuum-level cargo motion has been altered from that used by Williams \textit{et al.} \cite{Williams2021} to describe the motion of passive particles in active-passive systems.

First consider a general Langevin equation of the form \cite{Williams2021},

\begin{equation}
x(t+dt) = \mathcal{B}(x(t) + \Delta(dt) + \sqrt{2D} \, w(dt)),
\label{eq:GeneralLangevin}
\end{equation}

\noindent where $dt$ is a small increment in time, $\Delta(dt)$ is the stochastic displacement due to Poisson processes within the time $dt$, and $w(dt)$ is the stochastic displacement due to a diffusive Wiener process within a time $dt$. The time dependent position $x(t)\equiv x_a(t)$ from the main text. The function $\mathcal{B}(x)$ applies boundary conditions to the system, which for this derivation will be periodic and of the form,

\begin{equation}
\mathcal{B}(x) = x - 2lh \qquad \textnormal{for } \qquad x \in [(2l-1)h, (2l+1)h],
\label{eq:PeriodicBoundaries}
\end{equation}

\noindent where the periodic domain of interest is $x \in [-h, h]$ (corresponding to $n=0$). In the case of cargo motion $\Delta(dt)$ describes displacements due to stochastic binding or unbinding events. The stochastic Brownian displacement term $\sqrt{2D} \, w(dt)=\sqrt{2D} \, (W(t+dt)-W(t))=\sqrt{2D} \int_t^{t+dt}dW(t')\sim \mathcal{N}(0, 2D dt)$ (where $\mathcal{N}(0, 2D dt)$ is the normal distribution with mean $0$ and variance $2Ddt$) has been included for completeness, but is not used in this work \cite{Risken1989, VanKampen1992, Paul2013}. As a result of the boundary conditions defined in eq.(\ref{eq:GeneralLangevin} \& \ref{eq:PeriodicBoundaries}), it can be assumed that $P(x,t)=P(x-2lh,t)$ ($l \in \mathbb{Z}$), where $P(x,t)$ describes the probability of a cargo being at the position $x$ at time $t$.

In order to derive the evolution of the PDF $P(x,t+dt) = \langle\delta(x-x(t+dt)) \rangle$ (averaged over realisations of noise), the PDFs describing the dynamics of the Poisson and Wiener process terms in eq.(\ref{eq:GeneralLangevin}) must be defined \cite{Williams2021}. In this work, the PDF describing the probability of the Poisson process generating a displacement $\Delta(dt)=x_J$ from the position $x'$ within a time $dt$ has been defined as,

\begin{equation}
\begin{aligned}
p_{dt}(x_J|x') & \simeq \left( 1-\sum\limits_{n=1}^N \left[ P_n(x') \sum\limits_{m=1}^M \left( \bar{k}_m(x',n) \right) \right] dt \right) \delta(x_J) \\
& \quad\,\, + \sum\limits_{n=1}^N \left[ P_n(x') \sum\limits_{m=1}^M \left( \bar{k}_m(x',n)\, q_m(x_J|x',n) \right) \right] dt +\mathcal{O}(dt^2) \\
& = (1-S_k(x')dt)\,\delta(x_J)+S_q(x_J|x')dt + \mathcal{O}(dt^2),
\end{aligned}
\label{eq:MoveProb}
\end{equation}

\noindent where $\bar{k}_m(x',n)$ is the position dependent rate of the $m^\textnormal{th}$ type of event occurring when the cargo has $n$ legs bound (see eq.(\ref{eq:DiscreteRate})), and $q_m(x_J|x',n)$ is the corresponding probability of this event resulting in the displacement $x_J$. For comparison, this means that eq.(\ref{eq:DiscreteMoment}) can be rewritten $\lambda_m^{(i)}(x,n) = \int_{-\infty}^{+\infty} dx_J\, (x_J^i \,q_m(x_J|x,n))$, such that $S_\lambda^{(i)}(x)=\int_{-\infty}^{\infty} dx_J\,(x_J^i \,S_q(x_J|x))$. The first term in eq.(\ref{eq:MoveProb}) describes the probability of no displacement occurring ($x_J=0$), and the second term describes the probability of a non-zero displacement occurring. The form of eq.(\ref{eq:MoveProb}) deviates from that used by Williams \textit{et al.} \cite{Williams2021} as a result of introducing $N$ possible cargo states that can each exhibit $M$ different types of event.

The PDF $r_{dt}(\eta)=(1/\sqrt{4\pi D dt})\textnormal{exp}(-\eta^2 / 4Ddt) \sim \mathcal{N}(0,2Ddt)$ describes the probability of the Wiener process resulting in a displacement $\sqrt{2D}\,w(dt)=\eta$ within a time $dt$ \cite{Williams2021}. This means that the Wiener process term is independent of cargo position.

The analytical form of $P(x,t+dt)$ can be derived using the chosen distributions for the Poisson and Wiener processes such that,

\begin{equation}
\begin{aligned}
P(x,t+dt) & = \left\langle \delta(x - \mathcal{B}(x(t) + x_J + \eta)) \right\rangle \\
& = \int\limits_{-h}^{+h} dx' \, P(x',t) \int\limits_{-\infty}^{+\infty} dx_J \, p_{dt}(x_J|x') \int\limits_{-\infty}^{+\infty} d\eta \, r_{dt}(\eta) \\ 
& \qquad\qquad\qquad\qquad\qquad\qquad\,\, \times \sum\limits_{l=-\infty}^{+\infty} \delta(x-(x'+x_J+\eta-2lh)) \\
& = \int\limits_{-h}^{+h} dx'\, P(x',t) \int\limits_{-\infty}^{+\infty} dx_J\, p_{dt}(x_J|x') \left[ \sum\limits_{l=-\infty}^{+\infty} \left( \frac{e^{-\frac{(x-(x'+x_J-2lh))^2}{4Ddt}}}{\sqrt{4\pi Ddt}} \right) \right],
\end{aligned}
\label{eq:PeriodicFPDerivation}
\end{equation}

\noindent where the sum over $l$ enforces the periodicity of the system by stating that cargo initially at a position within the domain $x' \in [-h,h]$ contribute to the probability $P(x,t+dt)$ if they are at any position $x-2lh$ ($l \in \mathbb{Z}$) following a displacement. In order to convert eq.(\ref{eq:PeriodicFPDerivation}) into a Fokker-Planck equation, it is necessary to calculate its Fourier Tranform with respect to the position $x$ noting that,

\begin{equation}
\frac{\partial \tilde{P}(k,t)}{\partial t} = \lim\limits_{dt \rightarrow 0} \left( \frac{\tilde{P}(k,t+dt) - \tilde{P}(k,t)}{dt} \right).
\label{eq:FourierTransformDerivation}
\end{equation}

\noindent Since the position $x$ in eq.(\ref{eq:PeriodicFPDerivation}) represents the final position of cargo, and it can be assumed that $P(x,t)=P(x-2lh,t)$ ($l \in \mathbb{Z}$), the Fourier transform in eq.(\ref{eq:FourierTransformDerivation}) can be calculated using an integral over all space. Further derivation also requires the definition of the two Fourier transform identities,

\begin{equation}
\begin{aligned}
\lim\limits_{dt \rightarrow 0} \left[ \int\limits_{-\infty}^{+\infty} dx\, e^{-ikx}\left( \frac{e^{-\frac{(x-(x'+x_J-2lh))^2}{4Ddt}}}{\sqrt{4\pi Ddt}} \right) \right] & = \lim\limits_{dt \rightarrow 0} \left( e^{-ik(x'+x_j-2lh)}e^{-k^2 Ddt} \right) \\
& \simeq e^{-ik(x'+x_j-2lh)}(1-k^2 Ddt+...),
\end{aligned}
\label{eq:FourierIdentity1}
\end{equation}

\begin{equation}
\begin{aligned}
& \int\limits_{-h}^{+h} dx'\,f(x') \int\limits_{-\infty}^{+\infty} \frac{dk}{2\pi}\, \sum\limits_{l=-\infty}^{+\infty} \left( e^{ik(x-x'+2lh)} \right) \\
& \qquad\qquad\qquad\qquad\qquad = \int\limits_{-h}^{+h} dx'\,f(x') \, \sum\limits_{l=-\infty}^{+\infty} \left( \delta(x-x'+2lh) \right) \\
& \qquad\qquad\qquad\qquad\qquad = \sum\limits_{l=-\infty}^{+\infty} \left( f(x+2lh) \right) \qquad \textnormal{for } x+2lh \in [-h,h] \\
& \qquad\qquad\qquad\qquad\qquad = f(x) \qquad \textnormal{for } x \in [-h,h],
\end{aligned}
\label{eq:FourierIdentity2}
\end{equation}

\noindent where terms of $\mathcal{O}(dt^a)$ with $a>1$ have been neglected in eq.(\ref{eq:FourierIdentity1}).

The Fokker-Planck equation governing the motion of cargo while bound to a microtubule can be derived by substituting eq.(\ref{eq:MoveProb}, \ref{eq:PeriodicFPDerivation}, \ref{eq:FourierIdentity1} \& \ref{eq:FourierIdentity2}) into eq.(\ref{eq:FourierTransformDerivation}), such that,

\begin{equation}
\begin{aligned}
\frac{\partial P(x,t)}{\partial t} & = D \frac{\partial^2 P(x,t)}{\partial x^2} - S_k(x)\,P(x,t) \\
& \qquad + \int\limits_{-h}^{+h} dx' \, P(x',t) \,  \sum\limits_{l=-\infty}^{+\infty} \left( S_q((x+2lh)-x'|x') \right).
\end{aligned}
\label{eq:PeriodicFP}
\end{equation}

\noindent The final term of eq.(\ref{eq:PeriodicFP}) states that there is a contribution to the probability $P(x,t)$ when a particle that is at a position $x' \in [-h,h]$ at time $t$ jumps to a position $x+2lh$ within a time $dt$. The Fokker-Planck equation defined in eq.(\ref{eq:PeriodicFP}) can alternatively be derived by assuming that cargo obey a langevin equation that does not include the effects of periodic boundaries ($\mathcal{B}(x)=x$) by enforcing periodicity after generating a Fokker-Planck equation for an infinite domain. As well as assuming $P(x,t)=P(x-2lh,t)$ ($l \in \mathbb{Z}$), this method also requires the assumptions that $\bar{k}_m(x,n)=\bar{k}_m(x-2lh,n)$ and $q_m((x+2lh)-x'|x',n)=q_m(x-(x'-2lh)|x'-2lh)$, such that $S_q((x+2lh)-x'|x')=S_q(x-(x'-2lh)|x'-2lh)$.

In the limit where cargo can only exhibit small displacements ($x_J = x-x' \ll h$) in a time $dt$, such that $q_m(x_J|x',n)$ decays quickly as a function of $x_J$, eq.(\ref{eq:PeriodicFP}) can be simplified to include only the $l=-1,0,+1$ terms of the infinite sum. Provided that the boundaries to the periodic domain at $x=\pm h$ are far from any fluctuations in the $S_q(x_J|x)$ distribution away from zero, and that the characteristic unbinding timescales of cargo are much smaller than the average time it would take for them to cross the domain, the solution to a simplified Fokker-Planck equation in the limit $h \rightarrow \infty$ including only the $l=0$ term of eq.(\ref{eq:PeriodicFP}) will be a good approximation to the solution of the complete equation. This is equivalent to neglecting the periodicity of the system. In this case, a Kramers-Moyal expansion can be used to simplify the third term of eq.(\ref{eq:PeriodicFP}) by defining $x'=x-x_J$, and assuming that the displacements $x_J$ due to binding or unbinding events are small \cite{Risken1989}. This results in the recognisable Fokker-Planck equation \cite{Williams2021},

 \begin{equation}
\frac{\partial P(x,t)}{\partial t} = \frac{\partial}{\partial x} \left[ D_{eff}(x) \frac{\partial P(x,t)}{\partial x} \right] - \frac{\partial}{\partial x} \left[ v_{eff}(x) P(x,t) \right],
\label{eq:FP2}
\end{equation}

\noindent with the position dependent effective velocity ($v_{eff}(x)$) and diffusivity ($D_{eff}(x)$) terms defined in eq.(\ref{eq:Veff} \& \ref{eq:Deff}) in the main text (where $S_\lambda^{(i)}(x) = \int_{-\infty}^{+\infty} dx_J \,(x_J^i\, S_q(x_J|x))$ since $\lambda_m^{(i)}(x,n) = \int_{-\infty}^{+\infty} dx_J \,(x_J^i\, q_m(x_J|x,n))$). The Fokker-Planck equation defined in eq.(\ref{eq:FokkerPlanck}) in the main text requires the addition of terms associated with binding dynamics to eq.(\ref{eq:FP2}).

The effective velocity can be easily obtained from simulations as $S_\lambda^{(1)}(x)=k_t(x)\,\lambda^{(1)}(x)$, using the position dependent total rate ($k_t(x)$) and average displacement ($\lambda^{(1)}(x)$) of any event occurring. This can be shown to be equivalent to eq.(\ref{eq:Veff}) since,

\begin{equation}
	\begin{aligned}
    S_\lambda^{(1)}(x) & = k_t(x)\,\lambda^{(1)}(x) \\
    & = k_t(x) \sum_{n=1}^{N} P_n(x) \left[ \sum_{m=1}^{M} P_m(x,n)\,\lambda_m^{(1)}(x,n) \right] \\
    & = \sum_{n=1}^{N} P_n(x) \left[ \sum_{m=1}^{M} k_m(x,n)\,\lambda_m^{(1)}(x,n) \right].
    \end{aligned}
\end{equation}

\noindent where $P_m(x,n) = k_m(x,n)/k_t(x)$ is the probability of the $m^\textnormal{th}$ type of event occurring when a cargo has $n$ legs bound. An equivalent equation can be derived for the effective diffusivity $D_{eff}(x)=k_t(x)\,\lambda^{(2)}(x)$.

\subsection{The Probability of Cargo Having $n$ Legs Bound}

The probability distribution $P_n(x)$ has been previously derived for cargo that can rebind from the $n=0$ state \cite{VanKampen1992, Klumpp2005}. Using these previously published formulae \cite{VanKampen1992, Klumpp2005}, $P_n(x)$ has been defined in this work by the distributions,

\begin{equation}
\begin{aligned}
P_n(x) & = \left( \frac{P_0(x)}{1-P_0(x)} \right) \prod\limits_{i=0}^{n-1} \left( \frac{\bar{k}_1(x,n)}{\bar{k}_2(x,n+1)} \right), \\
P_0(x) & = \left( 1 +  \sum\limits_{n=0}^{N-1} \prod\limits_{i=0}^n \left( \frac{\bar{k}_1(x,n)}{\bar{k}_2(x,n+1)} \right) \right)^{-1}.
\end{aligned}
\label{eq:Pn}
\end{equation}

\begin{figure*}[h!]
\centering
	\includegraphics[width=0.75\linewidth]{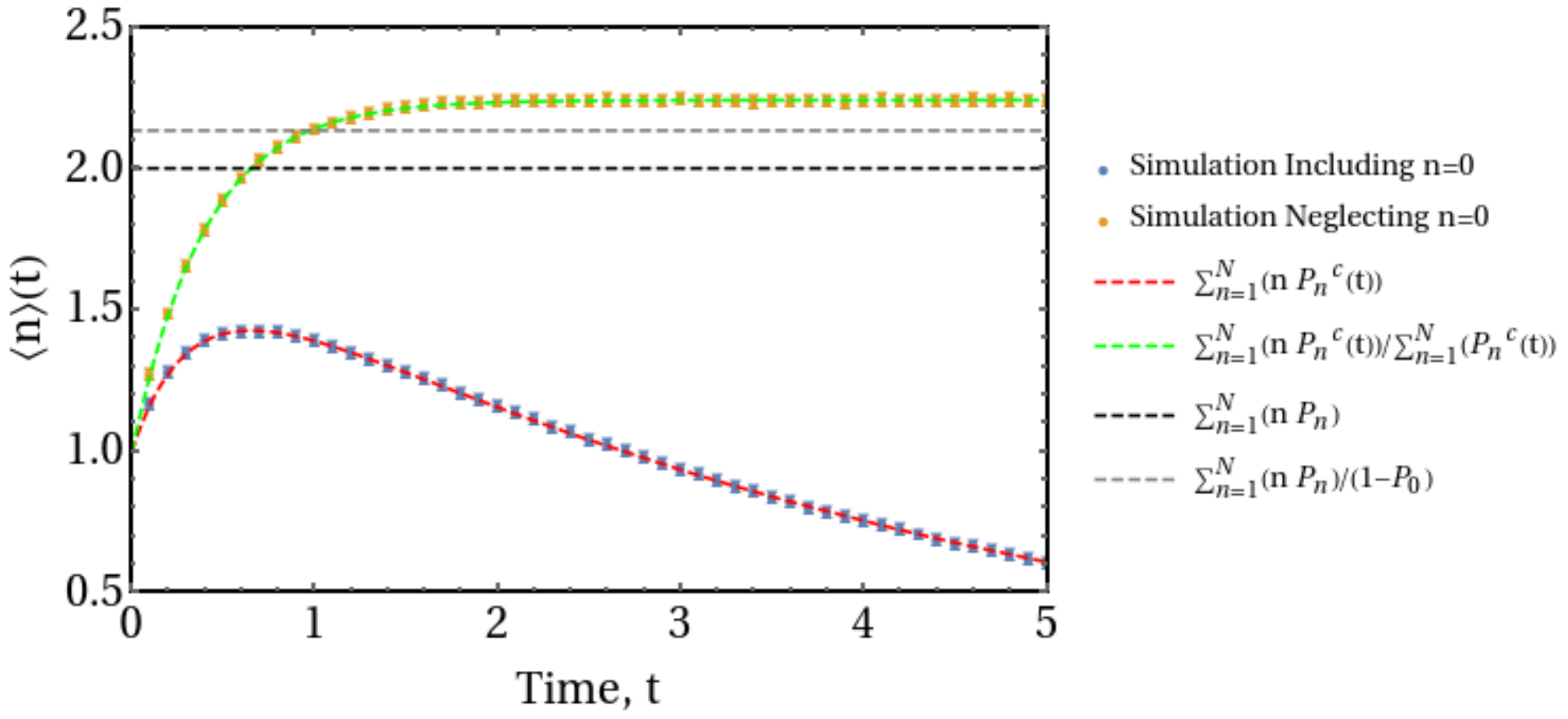}
 	\caption{Comparison of the evolution of the average number of bound legs $\langle n \rangle(t)$ for $4$-legged cargo obtained from simulations to various analytical approximations.}
 	\label{fig:AvN}
\end{figure*}

\noindent Allowing rebinding skews the average number of bound legs $\langle n \rangle(x)$ observed for bound cargo towards smaller values, which increases the magnitudes of $v_{eff}(x)$ and $D_{eff}(x)$ calculated using eq.(\ref{eq:Veff} \& \ref{eq:Deff}). In reality, $P_n(x,t)$ is a time dependent distribution that generates a time dependent distribution $\langle n \rangle(x,t)$. This is shown analytically and using simulations for a system with position independent binding and unbinding rates in Fig.(\ref{fig:AvN}). Complete calculations of $v_{eff}(x)$ and $D_{eff}(x)$ would therefore require an additional average over a cargo's dwell time distribution.

The complete analytical treatment of $\langle n \rangle(t)$ shown in Fig.(\ref{fig:AvN}) was carried out by first defining the set of linear ordinary differential equations that describe the binding dynamics of cargo legs. This set can be written as,

\begin{equation}
\begin{aligned}
\frac{d}{dt}
\begin{pmatrix}
P_1(t) \\
P_2(t) \\
\vdots \\
P_N(t)
\end{pmatrix} & =
\begin{pmatrix}
-k_{1,0} - k_{1,2} & k_{2,1} & 0 & \dots & 0 \\
k_{1,2} & -k_{2,1}-k_{2,3} & k_{3,2} & \dots & 0 \\
0 & k_{2,3} & -k_{3,2}-k_{3,4} & \dots & 0 \\
\vdots & \vdots & \vdots & \ddots & \vdots \\
0 & 0 & 0 & \dots & -k_{N,N-1} \\
\end{pmatrix} 
\begin{pmatrix}
P_1(t) \\
P_2(t) \\
\vdots \\
P_N(t)
\end{pmatrix} \\
& = \underline{\underline{\kappa}}\,\underline{P}(t),
\end{aligned}
\label{eq:TransitionMatrix}
\end{equation}

\noindent where the rates $k_{X,Y}$ represent the transition rates from the bound state $X$ to the bound state $Y$ (such that $k_{n,n+1}\equiv \bar{k}_{1}(n)$ and $k_{n+1,n}\equiv \bar{k}_{2}(n+1)$), $\underline{\underline{\kappa}}$ is the corresponding transition matrix, and $\underline{P}(t)$ is a state vector. The binding of new cargo is not included in eq.(\ref{eq:TransitionMatrix}) so that $t$ is a measure of the time since cargo first bound to the microtubule. Since cargo are assumed to always bind in the $n=1$ state, $\underline{P}(t)=(1,0,\cdots,0)$.

The eigen-values of $\underline{\underline{\kappa}}$ in eq.(\ref{eq:TransitionMatrix}), $e_n$, will ultimately dictate the characteristic decay rates of $P_n(t)$. In the case where all $k_{X,Y} > 0$ the transition matrix $\underline{\underline{\kappa}}$ is diagonalisable and can be written $\underline{\underline{\kappa}}=\underline{\underline{S}}\,\underline{\underline{D}}\,\underline{\underline{S}}^{-1}$, where $\underline{\underline{D}}$ contains only the eigen-values of $\underline{\underline{\kappa}}$ along its diagonal and $\underline{\underline{S}}$ consists of only the eigen-vectors corresponding to those eigen-values. Using these definitions, the solution of eq.(\ref{eq:TransitionMatrix}) can be derived as equal to,

\begin{equation}
\begin{aligned}
\underline{P}(t) & = \exp\left( \underline{\underline{\kappa}}\,t \right) \underline{P}(0) \\
& = \left( 1 + \underline{\underline{\kappa}}\, t + \frac{(\underline{\underline{\kappa}}\, t)^2}{2} + ... \right)\underline{P}_0(x) \\
& = \underline{\underline{S}} \left( 1 + \underline{\underline{D}}\, t + \frac{(\underline{\underline{D}}\, t)^2}{2} + ... \right) \underline{\underline{S}}^{-1} \underline{P}_0(x) \\
& =\underline{\underline{S}} \begin{pmatrix}
\textnormal{exp}(e_1\, t) & 0 & 0 & \cdots & 0 \\
0 & \textnormal{exp}(e_2\, t) & 0 & \cdots & 0 \\
0 & 0 & \textnormal{exp}(e_3\, t) & \cdots & 0 \\
\vdots & \vdots & \vdots &  \ddots & \vdots \\
0 & 0 & 0 & \dots & \textnormal{exp}(e_N\, t) \\
\end{pmatrix} \underline{\underline{S}}^{-1} \underline{P}_0(x).
\end{aligned}
\label{eq:PnDerivation}
\end{equation}

\noindent The result of eq.(\ref{eq:PnDerivation}) can be verified by using it to calculate the dwell time distribution and average dwell times of cargo, defined (respectively) as,

\begin{equation}
t_{dwell}(N,t) = -\frac{d}{dt}\left( \sum\limits_{n=1}^N P_n(t) \right),
\label{eq:TDwellDist}
\end{equation}

\begin{equation}
\langle t_{dwell} \rangle(N) = \int\limits_{0}^{\infty} dt\,\left( t_{dwell}(N,t)\,t \right),
\label{eq:AvTDwell}
\end{equation}

\begin{figure*}[h!]
\centering
	\includegraphics[width=1.0\linewidth]{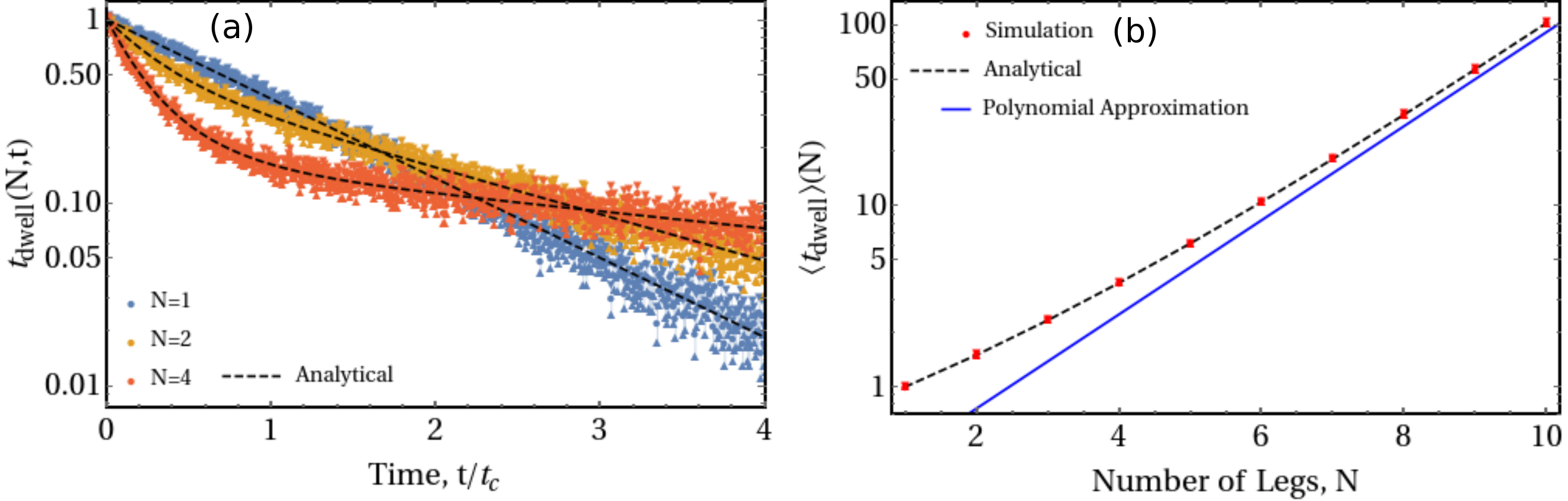}
 	\caption{(a) Cargo dwell time distributions for cargo with $N=1,2,4$, and (b) evolution of the average dwell time of cargo obtained from stochastic simulations ($n_{sim}$ same as in Fig.(\ref{fig:Diffusivity}b)) compared to values obtained using eq.(\ref{eq:TDwellDist} \& \ref{eq:AvTDwell}) respectively.}
 	\label{fig:TDwell}
\end{figure*}

\noindent Analytical distributions obtained using eq.(\ref{eq:TDwellDist} \& \ref{eq:AvTDwell}) have been shown to agree with those obtained from stochastic cargo simulations in Fig.(\ref{fig:TDwell}). The form of the dwell time distribution defined in eq.(\ref{eq:TDwellDist}) agrees qualitatively with that proposed by Klumpp \textit{et al.} \cite{Klumpp2005}, and results in multi-exponential distributions for cargo. The form of the average dwell time distribution defined in eq.(\ref{eq:AvTDwell}) tends towards the polynomial $\langle t_{dwell} \rangle(N) \sim \alpha\,\beta^N$ in the limit of large $N$ (where $\alpha$ is a fitting parameter), as shown in Fig.(\ref{fig:TDwell}).

In this work, the complete analysis detailed above has only been used to calculate the $N$-dependent evolution of the effective diffusivity in Fig.(\ref{fig:Diffusivity}b), and elsewhere the long-time behaviour of cargo has been probed by assuming that $P_n(x)$ defined in eq.(\ref{eq:Pn}) is a reasonable approximation for $P_n^c(x)$, defined as the steady-state value of $P_n(x,t)$ at times much greater than the cargo's average dwell time (see Fig.(\ref{fig:AvN})). Only bound cargo ($n>0$) contribute to the calculation of the distribution $P_n^c(x)$, since the probability of cargo being in the $n=0$ state should not affect the motion of bound cargo.

\subsection{Deriving the Langevin Equation Describing Cargo Motion}

In the absence of periodic boundary conditions, the general Langevin equation in eq.(\ref{eq:GeneralLangevin}) can be rewritten in the form \cite{Risken1989, Paul2013},

\begin{equation}
dx(t) = f(x(t))dt + g(x(t))dW(t),
\label{eq:IntroLangevin2}
\end{equation}

\noindent where $dx(t)$ is the infinitesimally small change in position that occurs within the time $dt$, and $\int_t^{t+dt}dW(t')=W(t+dt)-W(t)=w(dt)\sim \mathcal{N}(0, dt)$ \cite{Risken1989, VanKampen1992, Paul2013}. Other important relations involving the Wiener process include $w(0)=0$, $\langle w(t) \rangle = 0$, and $\langle w(t)w(t') \rangle = \textnormal{min}(t,t')$, where averages have been taken over realisations of noise \cite{Risken1989, VanKampen1992, Paul2013}.

Solving eq.(\ref{eq:IntroLangevin2}) requires integrating both sides of the equation between the times $t$ and $t+dt$, corresponding to the positions $x(t)=x_0$ and $x(t+dt)=x_0+dx(t)$ respectively. This can be achieved by assuming that the displacements $dx(t)$ are small and taking the Taylor expansion of $f(x(t))$ and $g(x(t))$, such that \cite{Risken1989},

\begin{equation}
\begin{aligned}
h(x(t')) & \simeq h(x_0) + \left. \frac{\partial h(x(t'))}{\partial x(t')} \right\vert_{x_0} dx(t') + ... \\
& \simeq h(x_0) + h'(x_0)(x(t')-x_0),
\end{aligned}
\label{eq:IntroKramersMoyal}
\end{equation}

\noindent where the second line truncates the series at $\mathcal{O}(dx(t'))$. Using eq.(\ref{eq:IntroKramersMoyal}), an integral of eq.(\ref{eq:IntroLangevin2}) can be solved iteratively up to $\mathcal{O}(dt)$ to give the moments of the displacement distribution \cite{Risken1989},

\begin{equation}
\langle x(t+dt) - x_0 \rangle = f(x_0) dt + g(x_0)g'(x_0)\left\langle \int_t^{t+dt} w(t') dW(t') \right\rangle,
\label{eq:ItoStratanovich1}
\end{equation}

\begin{equation}
\langle (x(t+dt) - x_0)^2 \rangle = (g(x_0))^2 dt.
\label{eq:ItoStratanovich2}
\end{equation}

The second term in eq.(\ref{eq:ItoStratanovich1}) cannot be solved using normal calculus methods, as it depends on the point in time at which the Wiener process term is evaluated. Discretising the second term in eq.(\ref{eq:ItoStratanovich1}) generates the equation \cite{Paul2013},

\begin{equation}
\left\langle \int_t^{t+dt} w(t') dW(t') \right\rangle = \left\langle \lim\limits_{\Delta \tau\rightarrow 0} \sum\limits_{i=0}^{N-1} \left[ w(\beta\,\tau_{i+1} + (1-\beta)\tau_i) \left( w(\tau_{i+1})-w(\tau_i) \right) \right] \right\rangle,
\label{eq:ItoStratanovich3}
\end{equation}

\noindent where $\Delta\tau=\tau_{i+1}-\tau_i$, $\tau_0 = t$, $\tau_N=t+dt$, and $\beta$ governs the time at which the Wiener process term is evaluated. Since the displacement of cargo has been defined as purely due to the binding dynamics of their legs in the lab frame, such that no future information about a cargo's position or configuration is required at the point of evaluation of the Wiener process term in eq.(\ref{eq:IntroLangevin2}), the {\^I}to convention ($\beta=0$) has been used to solve eq.(\ref{eq:ItoStratanovich3}) in this work. In this case, the average defined in eq.(\ref{eq:ItoStratanovich3}) is equal to zero, so eq.(\ref{eq:ItoStratanovich1}) becomes $\langle x(t+dt) - x_0 \rangle = f(x_0) dt$.

In order to derive how the results of eq.(\ref{eq:ItoStratanovich1} \& \ref{eq:ItoStratanovich2}) relate to the effective velocity and diffusivity defined in eq.(\ref{eq:Veff} \& \ref{eq:Deff}), the Chapman-Kolmogorov equation can be used to define the probability $P(x,t)$ of a cargo being at position $x$ at time $t$, and this can then be compared to eq.(\ref{eq:FokkerPlanck}) by inspection. The Chapman-Kolmogorov equation states \cite{Risken1989, VanKampen1992, Paul2013},

\begin{equation}
P(x,t+dt) = \int_{-\infty}^{\infty} (Q(x,t+dt|x',t)\,P(x',t))\, dx'
\end{equation}

\noindent where $Q(x,t+dt|x',t)$ is the transition probability of the particle moving from position $x'$ to position $x$ in a time $dt$. This equation can be re-arranged by defining the small displacement $\Delta=x-x'$ to be of the form \cite{Risken1989, VanKampen1992, Paul2013},

\begin{equation}
\begin{aligned}
P(x,t+dt) & = \int_{-\infty}^{\infty} \left[ \sum\limits_{n=0}^{\infty} \left( \left( \frac{(-\Delta)^n}{n!} \right) \frac{\partial^n}{\partial x^n} \left( Q(x+\Delta,t+dt|x,t) P(x,t) \right) \right) \right] d\Delta \\
& = \sum\limits_{n=0}^{\infty} \left[ (-1)^n \frac{\partial^n}{\partial x^n} \left( \left( \frac{M_n(x,t,dt)}{n!} \right) P(x,t) \right) \right],
\end{aligned}
\label{eq:IntroFPDerivation}
\end{equation}

\noindent where $M_n(x,t,dt)=\int_{-\infty}^{\infty}(\Delta^n \, Q(x+\Delta,t+dt|x,t))d\Delta$ are the moments of the displacement distribution. By truncating the series in eq.(\ref{eq:IntroFPDerivation}) at its third term, and taking the limit $dt\rightarrow 0$, a general Fokker-Planck equation can be defined that describes the time evolution of $P(x,t)$ \cite{Risken1989, VanKampen1992, Paul2013},

\begin{equation}
\frac{\partial P(x,t)}{\partial t} = -\frac{\partial}{\partial x} \left( V(x,t) P(x,t) \right) + \frac{\partial^2}{\partial x^2} \left( D(x,t) P(x,t) \right),
\label{eq:IntroFP}
\end{equation}

\noindent where $M_1(x,t,dt)\simeq V(x,t)\, dt$ and $M_2(x,t,dt)\simeq 2 D(x,t)\, dt$ are the net velocity and diffusivity terms respectively. These terms also correspond to the first and second moments of the displacement distribution respectively. Comparing eq.(\ref{eq:FokkerPlanck} \& \ref{eq:IntroFP}) and using the relations defined in eq.(\ref{eq:ItoStratanovich1} \& \ref{eq:ItoStratanovich2}) it can be observed that,

\begin{equation}
f(x(t)) = v_{eff}(x(t)) + \left. \frac{\partial D_{eff}(x)}{\partial x} \right\rvert_{x=x(t)} = S_\lambda^{(1)}(x(t)),
\end{equation}

\begin{equation}
g(x(t)) = \sqrt{2\,D_{eff}(x(t))} = \sqrt{S_\lambda^{(2)}(x(t))}.
\end{equation}

\subsection{Fitting the Bound Leg Distribution}

\begin{figure*}[h!]
\centering
	\includegraphics[width=1.0\linewidth]{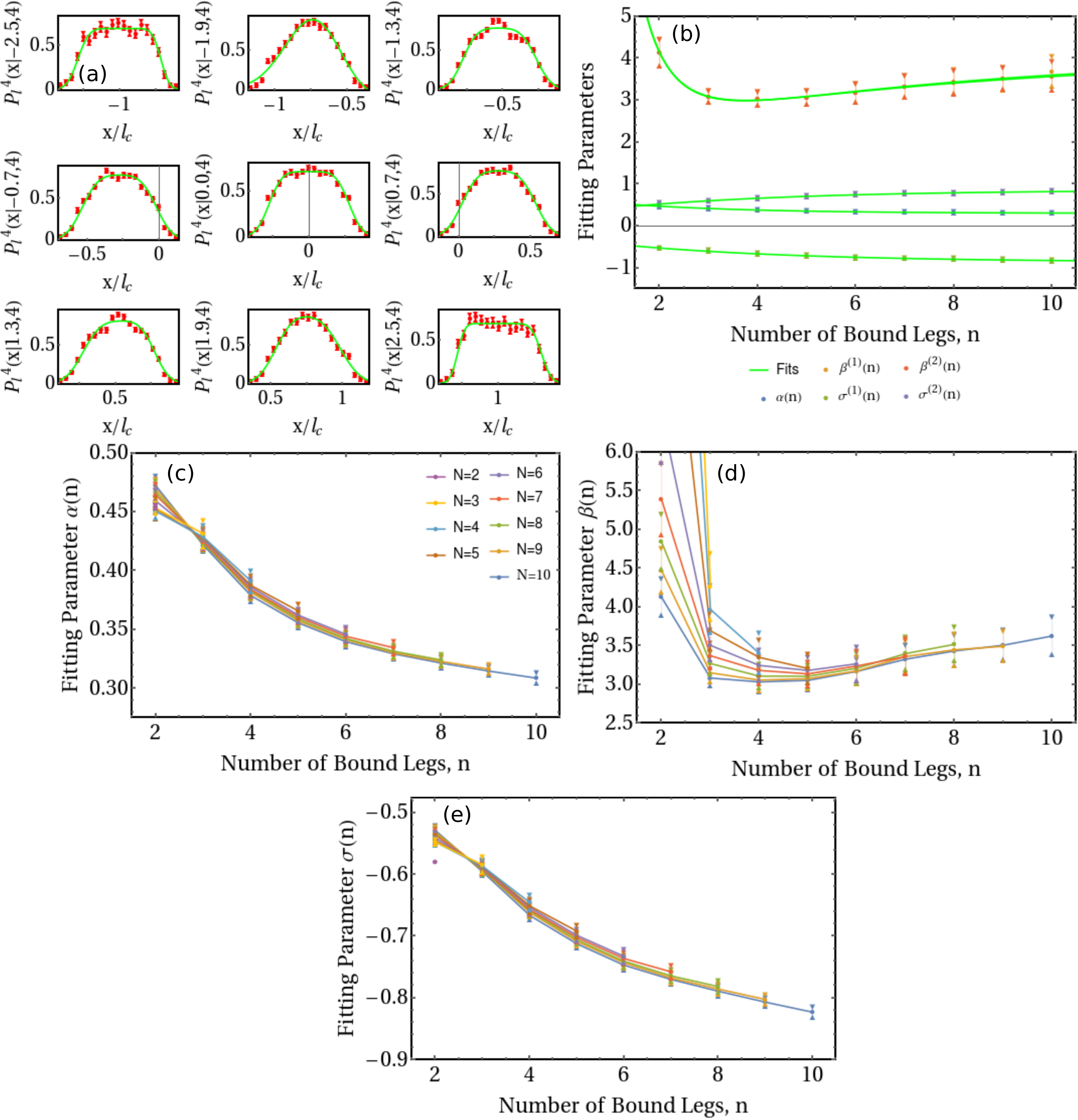}
 	\caption{(a) Example bound leg distributions (red) for $4$-legged cargo at different cargo centre positions (number of simulated cargo $n_{sim}=250\,000$). Fits (green) were calculated using eq.(\ref{eq:BoundLegDistribution}). Asymmetry can be observed most easily in the $P_l^4(x|1.9,4)$ and $P_l^4(x|-1.9,4)$ distributions, corresponding to positions near where $k_1(x)$ exhibits a maximum gradient (see Fig.(\ref{fig:BindingRate})). (b) Fits (green) to the fitting parameters obtained by fitting the bound leg distributions in Fig.(\ref{fig:Diffusivity}a) with eq.(\ref{eq:BoundLegDistribution}). (c,d,e) $N$-dependent evolution of the fitting parameters (c) $\alpha(n)$, (d) $\beta(n)$, and (e) $\sigma(n)$ defined in eq.(\ref{eq:BoundLegDistribution2}) ($n_{sim}$ same as in Fig.(\ref{fig:Diffusivity}b).}
 	\label{fig:BoundLegDist}
\end{figure*}

\begin{figure*}[h!]
\centering
	\includegraphics[width=0.5\linewidth]{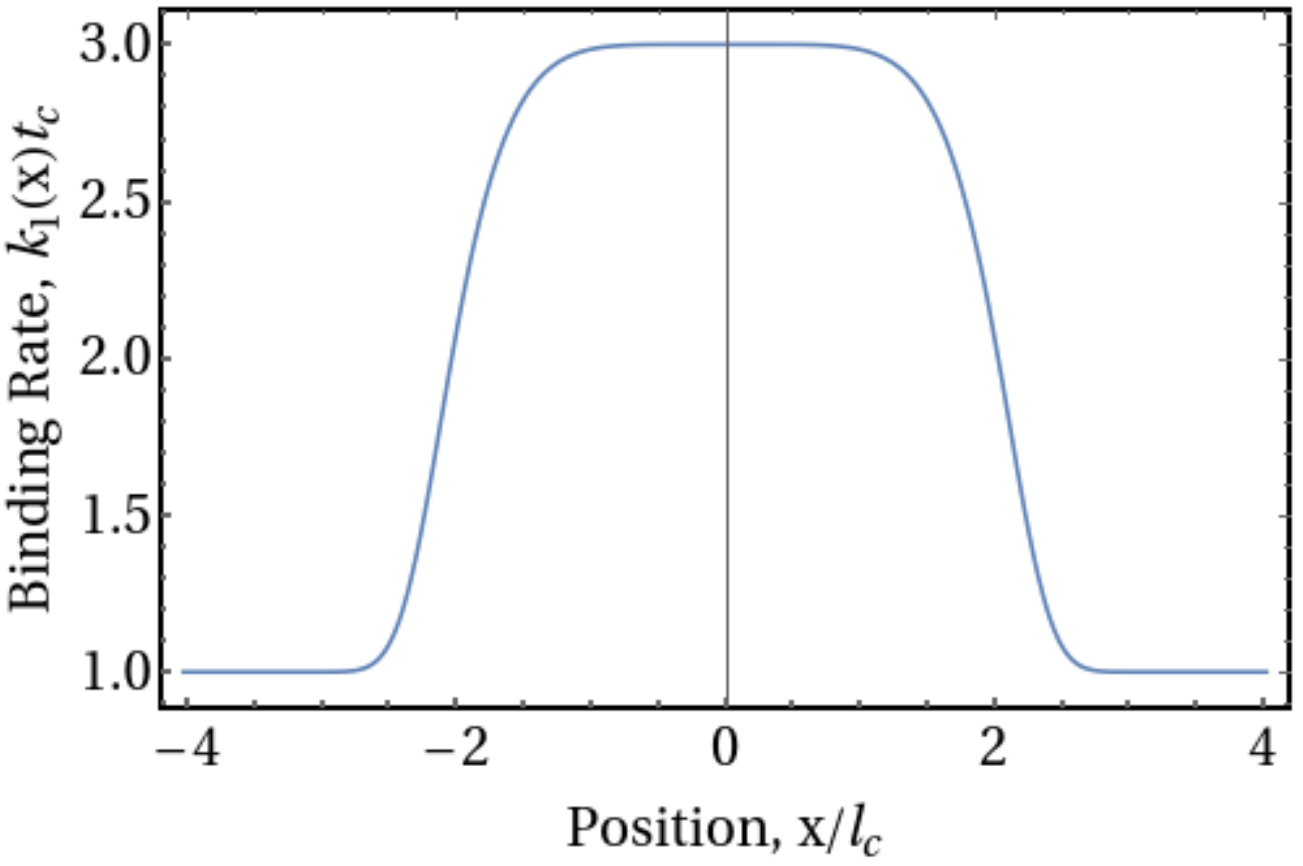}
 	\caption{The binding rate distribution defined in eq.(\ref{eq:PosDepRates}), which exhibits a central high cargo-substrate binding rate region and an outer region of constant binding rate.}
 	\label{fig:BindingRate}
\end{figure*}

The bound leg distributions $P_l^N(x|y,n)$ shown in this work (see Fig.(\ref{fig:Diffusivity}a \& \ref{fig:BoundLegDist}a)) were fitted using the equation,

\begin{equation}
F_l^N(x|y,n) = \alpha(y,n) \left( \textnormal{erf}\left(\beta^{(1)}(y,n)\left(x-\sigma^{(1)}(y,n)\right)\right) + \textnormal{erf}\left(\beta^{(2)}(y,n)\left(x-\sigma^{(2)}(y,n)\right)\right) \right),
\label{eq:BoundLegDistribution}
\end{equation}

\noindent where $\alpha(y,n)$, $\beta^{(1,2)}(y,n)$ and $\sigma^{(1,2)}(y,n)$ are fitting parameters. Asymmetry in $P_l^N(x|y,n)$ results in a non-zero component of the effective velocity due to unbinding events, since the moments defined in eq.(\ref{eq:DiscreteMoment}) become non-zero. This can be observed in Fig.(\ref{fig:BoundLegDist}a) for cargo that exhibit the binding rate distribution shown in Fig.(\ref{fig:BindingRate}). This asymmetry arises due to non-zero values of $\beta^{(1)}(y,n)-\beta^{(2)}(y,n)$ or $\sigma^{(1)}(y,n)+\sigma^{(2)}(y,n)$ for the fitting parameters defined in eq.(\ref{eq:BoundLegDistribution}). This phenomenon can be observed in Fig.(\ref{fig:BoundLegDist}a) near the positions of maximum gradient in the binding rate distribution (see Fig.(\ref{fig:BindingRate})).

The bound leg distributions of cargo must always be symmetric in the case of position independent binding and unbinding rates (when $n_{sim} \gg 1$). In this case, the bound leg distributions obtained from simulations can be averaged over all positions to minimise error, such that $P_l^N(x|n)=(\sum_{y=-h}^{y=+h} \, P_l^N(x|y,n))/(\sum_{x=b_l}^{x=b_u} \sum_{y=-h}^{y=+h} \, P_l^N(x|y,n) \Delta x)$ following normalisation using the discrete simulation grid spacing $\Delta x$ and the upper and lower bounds $b_{u,l}$ set by the shape factor (see Fig.(\ref{fig:Diffusivity}a)). In this case, the fit defined in eq.(\ref{eq:BoundLegDistribution}) can also be simplified to the form,

\begin{equation}
F_l^N(x|n) = \alpha(n) \left( \textnormal{erf}\left(\beta(n)\left(x-\sigma(n)\right)\right) + \textnormal{erf}\left(-\beta(n)\left(x+\sigma(n)\right)\right) \right),
\label{eq:BoundLegDistribution2}
\end{equation}

\noindent which contains fewer fitting parameters. This result is evidenced in Fig.(\ref{fig:BoundLegDist}b), since $\beta^{(1)}(n)\simeq\beta^{(2)}(n)$ and $\sigma^{(1)}(n)\simeq-\sigma^{(2)}(n)$ $\forall n$ after fitting $P_l^{10}(x|n)$ with eq.(\ref{eq:BoundLegDistribution}).

The evolution of the fit parameters $\alpha(n)$, $\beta(n)$ and $\sigma(n)$ defined in eq.(\ref{eq:BoundLegDistribution2}) has been plotted in Fig.(\ref{fig:BoundLegDist}c-e) as a function of $n$ for cargo with different numbers of legs, and it can be observed that the parameters tend towards smooth distributions in the limit $N \gg 1$. The evolution of the individual fit parameters was fitted using the equations $\alpha(n)=a_1\,\exp(-b_1\,n)+c_1$, $\beta(n)=(a_2/n^2)-(b_2/n)+c_2$ and $\sigma(n)=a_3\,\exp(-b_3\,n)+c_3$, where $a_{1,2,3}$, $b_{1,2,3}$ and $c_{1,2,3}$ are additional fitting parameters.

\subsection{Deriving Experimental Input Parameters for Cargo Binding Simulations}

\begin{figure*}[h!]
\centering
	\includegraphics[width=\linewidth]{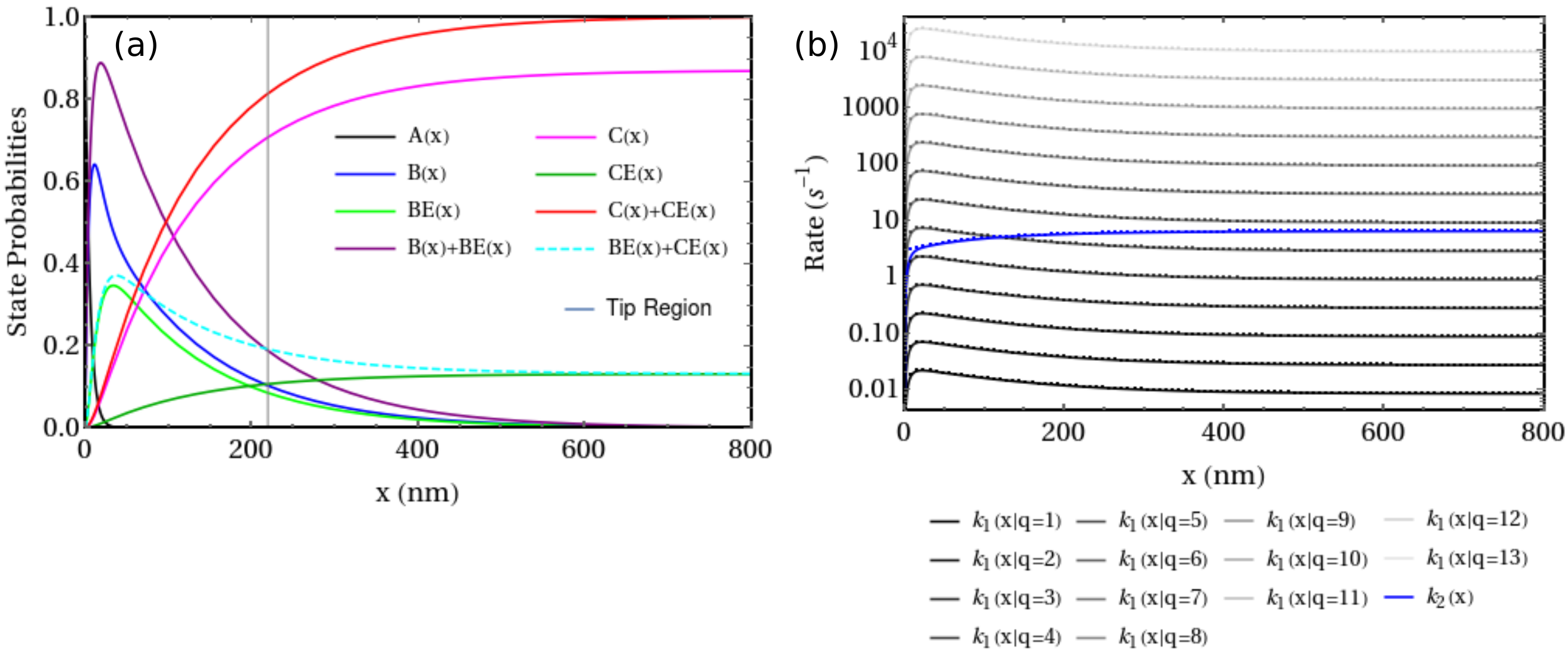}
 	\caption{(a) Steady-state probability distributions of tubulin heterodimer states as a function of position along a microtubule (the microtubule edge is set at $x_{edge}=0$ nm), obtained by substituting the experimental parameters from table \ref{tab:ExperimentalModelValues} into eq.(\ref{eq:TubulinStates2}). (b) Position dependent EB binding and unbinding rates obtained by substituting the experimental parameters from table \ref{tab:ModifiedModelValues} into eq.(\ref{eq:SingleEBRates}) (key below plot).}
 	\label{fig:ExpInputs}
\end{figure*}

Experimental parameters used as inputs for simulations were obtained from previously published work and are presented in table \ref{tab:ExperimentalModelValues}. These parameters were used to derive the position and time dependent probabilities of finding tubulin heterodimers in different states along microtubules by expanding upon the approach derived by Maurer \textit{et al.} \cite{Maurer2014}. In this work, lattice states have been added to the model to generate binding and unbinding rate distributions along entire microtubules. The model now includes the probabilities: $A(x,t)$ describing tubulin with an associated GTP molecule or that is part of the tapering microtubule end; $B(x,t)$ describing tubulin in the GDP-Pi state (the preferred binding site of EBs \cite{Roth2018}); $BE(x,t)$ describing tubulin in the GDP-Pi state and bound to an EB; $C(x,t)$ describing tubulin with an associated GDP molecule; $CE(x,t)$ describing tubulin with an associated GDP molecule and bound to an EB. For simplicity, it is assumed that EBs cannot bind to tubulin in the $A(x,t)$ state. The probabilities of finding tubulin in these different states in the rest frame of a microtubule end growing linearly with velocity $v_{MT}$ are the solutions to the set of equations,

\begin{equation}
    \begin{aligned}
    \frac{\partial A(x,t)}{\partial t} & = -v_{MT} \frac{\partial A(x,t)}{\partial x} - k_f\,A(x,t), \\
    \frac{\partial B(x,t)}{\partial t} & = -v_{MT} \frac{\partial B(x,t)}{\partial x} + k_f\,A(x,t) - (k_{on}^{tip}[EB]\Delta x+k_h)B(x,t) + k_{off}^{tip}\,BE(x,t), \\
    \frac{\partial BE(x,t)}{\partial t} & = -v_{MT} \frac{\partial BE(x,t)}{\partial x} + k_{on}^{tip}[EB]\Delta x\, B(x,t) - (k_{off}^{tip} + k_{EBh})BE(x,t),
    \end{aligned}
    \label{eq:TubulinStates1}
\end{equation}

\noindent with parameters defined in table \ref{tab:ExperimentalModelValues}. These equations assume non-competitive and non-cooperative EB binding to the microtubule. The steady-state solutions to the set of equations defined in eq.(\ref{eq:TubulinStates1}) are,

\begin{equation}
    \begin{aligned}
    A(x) &= A_0 \, e^{-\l_1 (x-x_0)}, \\
    B(x) &= B_0(e^{-l_1 (x-x_0)} + \gamma_1 \, e^{-l_2 (x-x_0)} - (1+\gamma_1) \, e^{-l_3 (x-x_0)}), \\
    BE(x) &= BE_0(e^{-l_1 (x-x_0)} + \gamma_2 \, e^{-l_2 (x-x_0)} - (1+\gamma_2) \, e^{-l_3 (x-x_0)}), \\
    C(x) &= \left( \frac{k_{off}^{lat}}{k_{on}^{lat}[EB]\Delta x+k_{off}^{lat}} \right) (1 - (A(x) + B(x) + BE(x))), \\
    CE(x) &= \left( \frac{k_{on}^{lat}[EB]}{k_{on}^{lat}[EB]\Delta x+k_{off}^{lat}} \right) (1 - (A(x) + B(x) + BE(x))),
    \end{aligned}
    \label{eq:TubulinStates2}
\end{equation}

\noindent where the characteristic length-scales $l_{1,2,3}$ and the relative amplitudes $\gamma_{1,2}$ are complicated functions of the transition rates between the states defined in eq.(\ref{eq:TubulinStates1} \& \ref{eq:TubulinStates2}). The distributions obtained by substituting the parameters from table \ref{tab:ExperimentalModelValues} into eq.(\ref{eq:TubulinStates2}) after applying the following corrections are shown in Fig.(\ref{fig:ExpInputs}a).

For use in simulations, the average binding rates and dwell times presented in table \ref{tab:ExperimentalModelValues} must first be corrected by taking into account the probability of a potential EB binding site being in the GDP-Pi or GDP-bound states. Assuming that $(B(x)+BE(x))/(C(x)+CE(x))\ll1$ on the lattice (see Fig.(\ref{fig:ExpInputs}a)), it can be shown iteratively that $(B(x)+BE(x))/(C(x)+CE(x)) \simeq 0.5$ within a distance $L_{tip}$ from the microtubule tip. This means that the correct dwell times and binding rates for EBs interacting with tubulin in the GDP-Pi state satisfy the equations $\tau^{tip}=0.5(\tau^{GDP-Pi}+\tau^{GDP})$ and $k_{on}^{tip}=0.5(k_{on}^{GDP-Pi}+k_{on}^{GDP})$ respectively (where $\tau^{GDP}\equiv \tau^{lat}$ and $k_{on}^{GDP}\equiv k_{on}^{lat}$ from table \ref{tab:ExperimentalModelValues}). Any co-operative interactions between EBs has been neglected in this work \cite{Lopez2014, Lopez2016, Zhang2018}.

Next, the binding rates must be converted from units of `per unit concentration of EBs in solution per unit length along the microtubule per unit time' to units of `per specific EB that is associated to a simulated cargo per unit time' using the equation,

\begin{equation}
    \kappa(x)(\textnormal{s}^{-1}) = \left( \frac{\Delta x [EB]}{N_{EB}^{av}} \right) k(x) (\textnormal{nM}^{-1}\mu\textnormal{m}^{-1}\textnormal{s}^{-1}) = \frac{k(x) (\textnormal{nM}^{-1}\mu\textnormal{m}^{-1}\textnormal{s}^{-1})}{\pi\,N_{A}((\rho_{MT}+\epsilon)^2-\rho_{MT}^2)},
    \label{eq:Conversion}
\end{equation}

\noindent where $N_{EB}^{av} = N_{A}[EB]V_{int}$ is the average number of EBs available to bind to the microtubule and $V_{int}=\pi((\rho_{MT}+\epsilon)^2-\rho_{MT}^2)\Delta x$ defines the cylindrical volume of EB-microtubule interactions with radius $\epsilon$. The radius $\epsilon$ is expected to be within the range $\rho_{EB} \leq \epsilon \leq \rho_D$, where $\rho_D = \sqrt{2\,D_{aq}^{EB}} \simeq 2\,700\,\rho_{EB}$ is the average distance an EB is expected to diffuse in solution in one second (see table \ref{tab:ExperimentalModelValues}). The resulting constant of proportionality that links $\kappa = \psi\, k$ must therefore be in the range $0.0036 \textnormal{ (mol}\, \mu\textnormal{m}^{-2}) \leq \psi \leq 4100\textnormal{ (mol}\, \mu\textnormal{m}^{-2})$ (where mol indicates the number of moles). The ranges of rates $\kappa$ for the microtubule tip and lattice are shown in table \ref{tab:ModifiedModelValues} and span approximately six orders of magnitude. In order to sample this parameter space efficiently the binding rates have been sampled uniformly in log-space, such that the rates used in simulations are,

\begin{equation}
\begin{aligned}
    k_1(x|q) & = \left[ \kappa_{on}^{GDP-Pi}\, (B(x)+BE(x)) + \kappa_{on}^{GDP}\, (C(x)+CE(x)) \right] \psi_{min} \left( \frac{\psi_{max}}{\psi_{min}} \right)^{\frac{q-1}{12}}, \\
    k_2(x) & = \frac{k_{off}^{GDP-Pi}(B(x)+BE(x)) + k_{off}^{GDP}(C(x)+CE(x))}{B(x)+BE(x)+C(x)+CE(x)},
    \end{aligned}
    \label{eq:SingleEBRates}
\end{equation}

\noindent where $\psi_{min,max}$ are defined as the limits of the range $\psi_{min} \leq \psi \leq \psi_{max}$, and $q \in [1, 13]$. The definitions in eq.(\ref{eq:SingleEBRates}) assume that the steady-state distributions of $B(x)$, $BE(x)$, $C(x)$, and $CE(x)$ are unchanged by the binding or unbinding of individual cargo legs, and neglect the effects of competition.

The distributions obtained by substituting the parameters from table \ref{tab:ModifiedModelValues} into eq.(\ref{eq:SingleEBRates}) are shown in Fig.(\ref{fig:ExpInputs}b). Simulations have been carried out using $q = 1, ... , 13$ across the phase-space $N \in [2,12]$ and $L \in [32, 192]$ nm, with representative results for $q=6$ shown in Fig.(\ref{fig:ExpSimulations}, \ref{fig:ExpBreakdown} \& \ref{fig:ExpDiffusivities}).

\subsection{Deriving the Fokker-Planck Equation Describing Cargo\\ Motion in $2$D}

The derivation of a $2$D Fokker-Planck equation for cargo motion follows that of the $1$D case for eq.(\ref{eq:MoveProb} \& \ref{eq:PeriodicFP}) with $x \rightarrow \underline{x}$ and $dx \rightarrow d\underline{x}$, although system periodicity and the Wiener process diffusion term are neglected in this case. A Kramers-Moyal expansion can again be used to simplify the derivation by assuming small $2$D displacements of magnitude $|\underline{x}_J|=|\underline{x}-\underline{x}'| \ll 1$. Together, these assumptions result in the equation,

\begin{equation}
\begin{aligned}
\frac{\partial P(\underline{x},t)}{\partial t} & = -S_k(\underline{x})\,P(\underline{x},t) + \int\limits_{\infty}^{+\infty} d\underline{x}' \, P(\underline{x}', t) S_q(\underline{x}_J|\underline{x}') \\
& = -S_k(\underline{x})\,P(\underline{x},t) + P(\underline{x}, t) \int\limits_{\infty}^{+\infty} d\underline{x}_J  \left( 1 - \underline{x}_J \cdot \underline{\nabla} + \left( \frac{(\underline{x}_J \cdot \underline{\nabla})^2}{2} \right) \right) S_q(\underline{x}_J|\underline{x}) \\
& = -\frac{\partial}{\partial x} \left[ P(\underline{x}, t) \int\limits_{\infty}^{+\infty} dx_J\, x_J \left( \int\limits_{\infty}^{+\infty} dy_J \, S_q(\underline{x}_J|\underline{x}) \right) \right] \\
& \qquad -\frac{\partial}{\partial y} \left[ P(\underline{x}, t) \int\limits_{\infty}^{+\infty} dy_J\, y_J \left( \int\limits_{\infty}^{+\infty} dx_J \, S_q(\underline{x}_J|\underline{x}) \right) \right] \\
& \qquad + \left( \frac{1}{2} \right) \frac{\partial^2}{\partial x^2} \left[ P(\underline{x}, t) \int\limits_{\infty}^{+\infty} dx_J\, x_J^2 \left( \int\limits_{\infty}^{+\infty} dy_J \, S_q(\underline{x}_J|\underline{x}) \right) \right] \\
& \qquad + \left( \frac{1}{2} \right) \frac{\partial^2}{\partial y^2} \left[ P(\underline{x}, t) \int\limits_{\infty}^{+\infty} dy_J\, y_J^2 \left( \int\limits_{\infty}^{+\infty} dx_J \, S_q(\underline{x}_J|\underline{x}) \right) \right] \\
& \qquad + \left( \frac{1}{2} \right) \frac{\partial^2}{\partial x\,\partial y} \left[ P(\underline{x}, t) \int\limits_{\infty}^{+\infty} dx_J\, x_J \left( \int\limits_{\infty}^{+\infty} dy_J \, y_J \, S_q(\underline{x}_J|\underline{x}) \right) \right], \\
& \equiv -\frac{\partial}{\partial x} \left( P(\underline{x}, t)\, S_\lambda^{(1,x)}(\underline{x}) \right) -\frac{\partial}{\partial y} \left( P(\underline{x}, t)\,S_\lambda^{(1,y)}(\underline{x}) \right) + \frac{\partial^2}{\partial x^2} \left( \frac{P(\underline{x}, t)\, S_\lambda^{(2,x)}(\underline{x})}{2} \right) \\
& \qquad + \frac{\partial^2}{\partial y^2} \left( \frac{P(\underline{x}, t)\, S_\lambda^{(2,y)}(\underline{x})}{2} \right) + \frac{\partial^2}{\partial x\,\partial y} \left( \frac{P(\underline{x}, t)\, S_\lambda^{(1,xy)}(\underline{x})}{2} \right)
\end{aligned}
\label{eq:2DFPDerivation}
\end{equation}

\noindent where $\int_{-\infty}^{+\infty} d\underline{x} \equiv \int_{-\infty}^{+\infty} \int_{-\infty}^{+\infty} dx\,dy$ and $S_\lambda^{(i,k)} = \int_{-\infty}^{+\infty} dx_J\, \int_{-\infty}^{+\infty} dy_J\, (k_J^i\,S_q(\underline{x_J}|\underline{x}))$. The Fokker-Planck equation defined in eq.(\ref{eq:2DFokkerPlanck}) in the main text is generated by rearranging the terms in the final line of eq.(\ref{eq:2DFPDerivation}) and splitting them up according to which generate displacements purely in the $x$ or $y$ directions.

Unlike for eq.(\ref{eq:DiscreteMoment} \& \ref{eq:DiscreteRate}) and the the $1$D cargo model, there is no general form for the $i^{th}$ moment of the cargo displacement distribution ($\lambda_m^{(i)}(\underline{x}_a(t),n)$) and the average rate ($\bar{k}_m(\underline{x}_a(t),n)$) associated with each type of event ($m=1,2$) in a $2$D system due to the presence of `cross' terms with both $x$ and $y$ direction dependence. However, individual formulae can be derived for these moments and rates of the forms,

\begin{equation}
\lambda_m^{(i,x)}(\underline{x}_a(t), n) = \left( \frac{1}{n +\Delta_m} \right)^i \left( \frac{\int\limits_{b_l^x}^{b_u^x} dx\, \int\limits_{b_l^y}^{b_u^y} dy\, \left([\Delta_m (x - x_a(t))]^i\, k_m(\underline{x}|\underline{x}_a(t),n) \right)}{\int\limits_{b_l^x}^{b_u^x} dx\, \int\limits_{b_l^y}^{b_u^y} dy\, k_m(\underline{x}|\underline{x}_a(t),n)} \right),
\label{eq:DiscreteMomentX}
\end{equation}

\begin{equation}
\lambda_m^{(i,y)}(\underline{x}_a(t), n) = \left( \frac{1}{n +\Delta_m} \right)^i \left( \frac{\int\limits_{b_l^x}^{b_u^x} dx\, \int\limits_{b_l^y}^{b_u^y} dy\, \left([\Delta_m (y - y_a(t))]^i\, k_m(\underline{x}|\underline{x}_a(t),n) \right)}{\int\limits_{b_l^x}^{b_u^x} dx\, \int\limits_{b_l^y}^{b_u^y} dy\, k_m(\underline{x}|\underline{x}_a(t),n)} \right),
\label{eq:DiscreteMomentY}
\end{equation}

\begin{equation}
\lambda_m^{(1,xy)}(\underline{x}_a(t), n) = \left( \frac{1}{n +\Delta_m} \right)^i \left( \frac{\int\limits_{b_l^x}^{b_u^x} dx\, \int\limits_{b_l^y}^{b_u^y} dy\, \left(\Delta_m (x - x_a(t))(y - y_a(t))\, k_m(\underline{x}|\underline{x}_a(t),n) \right)}{\int\limits_{b_l^x}^{b_u^x} dx\, \int\limits_{b_l^y}^{b_u^y} dy\, k_m(\underline{x}|\underline{x}_a(t),n)} \right),
\label{eq:DiscreteMomentXY}
\end{equation}

\begin{equation}
\bar{k}_m(\underline{x}_a(t), n) = \left( \frac{N\,\delta_{m,1}-n\,\Delta_m}{(b_u^x-b_l^x)(b_u^y-b_l^y)} \right) \int\limits_{b_l^x}^{b_u^x} dx\, \int\limits_{b_l^y}^{b_u^y} dy\, k_m(\underline{x}|\underline{x}_a(t),n),
\label{eq:2DDiscreteRate}
\end{equation}

\noindent where $\Delta_m=\delta_{m,1}-\delta_{m,2}$ (see eq.(\ref{eq:DiscreteMoment} \& \ref{eq:DiscreteRate})), $\delta_{i,j}$ is the Kronecker delta function, and $b_{u,l}^{x,y}$ are the upper and lower bounds of the averages in the $x$ and $y$ directions respectively, defined by the corresponding $2$D shape factor and bound leg distribution.

\clearpage

\section{Supplementary Methods}

\subsection*{Stochastic Cargo Binding Simulations}

The simulations used in this work follow the schematic in Fig.(\ref{fig:Model}), such that the dynamics of simulated cargo are a function of the number of legs they have available to bind ($N-n$, where $N$ is the total number of legs of the cargo and $n$ of these are currently bound), their width ($2L$), and their centre position ($x_a(t)$) at any time ($t$). It is assumed that simulated cargo exhibit the shape factor $S_0^N(x|x_a(t))$, so their legs can only bind at positions within the range $x_a(t)-L \leq x \leq x_a(t)+L$. In order to emulate the inter-tubulin distance along a microtubule, it is also assumed that cargo legs can only bind at discrete positions separated by the spacing $\Delta x$, although $x_a(t)=(1/n)\sum_{l=1}^{l=n}x_l$ is calculated as a continuous variable for cargo legs at positions in the set $\{ x_l \}(t)$. This means that cargo legs that exhibit the position dependent binding and unbinding rates $k_{1,2}(x)$ (respectively) also exhibit the position dependent total rates of a binding or unbinding event occurring $(N-n)\sum_{i=(x_a(t)-L)/\Delta x}^{i=(x_a(t)+L)/\Delta x}k_1(x_i)/(2(L/\Delta x)+1)$ and $\sum_{l=1}^{l=n} k_2(x_l)$ (respectively). Cargo exhibit the constant net velocity $\bar{k}_3\,\Delta x$, which has been implemented as a shift of magnitude $\Delta x$ with rate $\bar{k}_3$ in the positions of all of a cargo's bound legs.

\begin{figure*}[h!]
\centering
	\includegraphics[width=\linewidth]{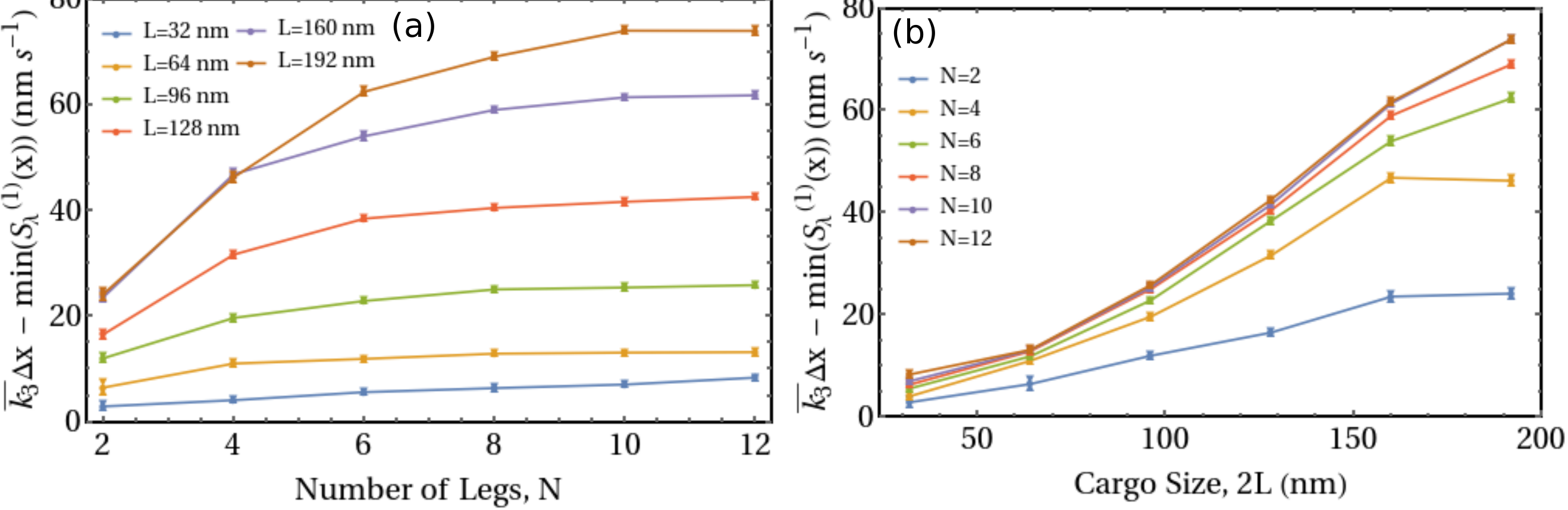}
 	\caption{Evolution of the maximum effective velocity resulting from only binding and unbinding events exhibited by cargo permanently bound to multiple EBs, obtained from stochastic cargo binding simulations ($n_{sim}$ the same as for Fig.(\ref{fig:ExpSimulations}b)) with (a) constant $L$ and varying $N$, and (b) constant $N$ and varying $L$. Maximum effective velocity increases monotonically before plateauing as a function of increasing $N$, and increases approximately quadratically with increasing $L$.}
 	\label{fig:ExpBreakdown}
\end{figure*}

\begin{figure*}[h!]
\centering
	\includegraphics[width=\linewidth]{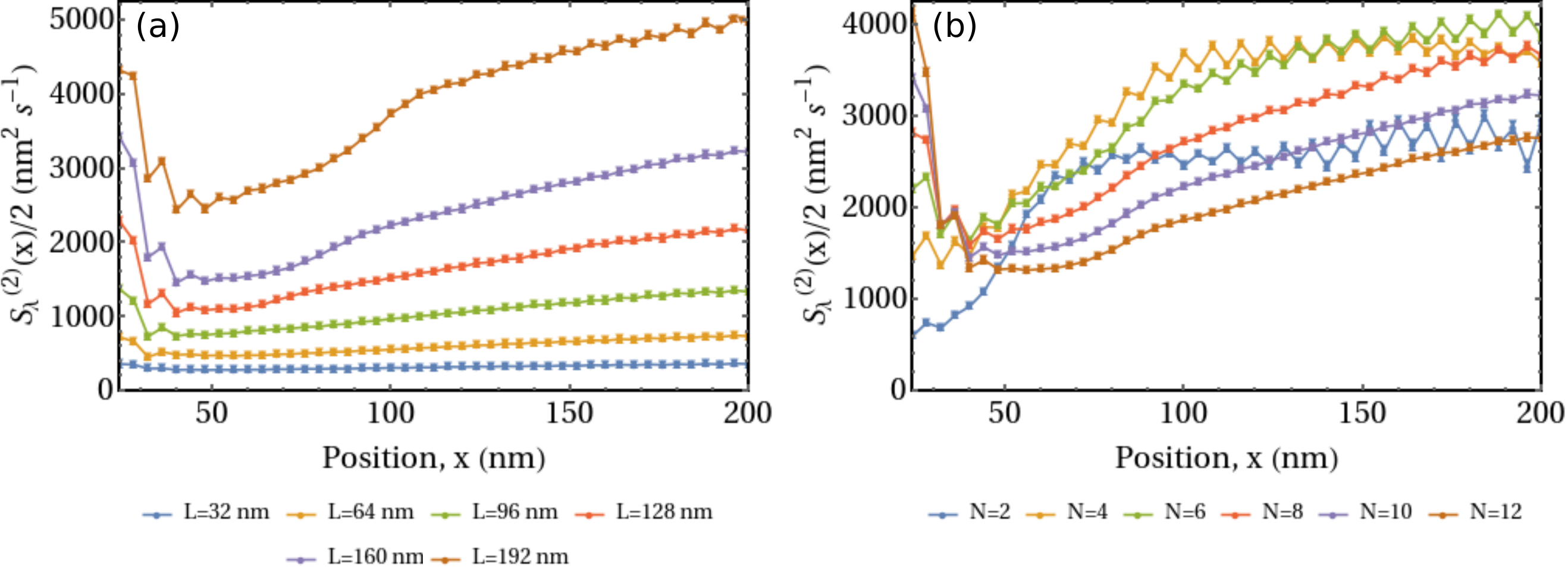}
 	\caption{The position dependent effective diffusivity exhibited by bound cargo (a) permanently bound to $N=10$ EBs ($n_{sim}=50\,000$), and (b) of size $2L=160$ nm ($n_{sim}=100\,000$ for $2 \leq N \leq 8$, $n_{sim}=50\,000$ for $N = 10$, and $n_{sim}=25\,000$ for $N = 12$), obtained from stochastic cargo binding simulations using input parameters derived from previously published experimental data presented in tables \ref{tab:ExperimentalModelValues} \& \ref{tab:ModifiedModelValues}. The microtubule edge is set at $x_{edge}=0$ nm. The position of the minimum of the effective diffusivity appears to be set by $N$, whereas its maximum magnitude increases monotonically as a function of $L$.}
 	\label{fig:ExpDiffusivities}
\end{figure*}

The simulations used to generate Fig.(\ref{fig:Diffusivity}, \ref{fig:Velocity}, \ref{fig:PDF}, \ref{fig:MSD}, \ref{fig:AvN}, \ref{fig:TDwell}, \ref{fig:BoundLegDist}, \ref{fig:VelocityComponents} \& \ref{fig:NonZeroVMT}) used $L/\Delta x=10$, a periodic domain of width $2h/L = h/(5\Delta x) = 40$, and the binding rate distribution defined in eq.(\ref{eq:PosDepRates}) (see Fig.(\ref{fig:BindingRate})). Of these simulations, only those used to generate Fig.(\ref{fig:PDF}a \& \ref{fig:NonZeroVMT}) used a non-zero net velocity of $\bar{k}_3\,\Delta x = 0.1\,L/t_c$. The simulations used to generate Fig.(\ref{fig:ExpSimulations}, \ref{fig:ExpBreakdown} \& \ref{fig:ExpDiffusivities}) used various values of $L$ (specified in the figure legends) and $\Delta x=8$~nm (see table \ref{tab:ExperimentalModelValues}). The binding and unbinding rate distributions used in these simulations were derived by substituting the parameters from tables \ref{tab:ExperimentalModelValues} \& \ref{tab:ModifiedModelValues} into eq.(\ref{eq:SingleEBRates}), and are shown in Fig.(\ref{fig:ExpInputs}).

Once all legs of a cargo are unbound ($n=0$), its dwell time is stored and the next cargo is simulated. This assumes that cargo are unlikely to rebind a leg in the time it takes for them to diffuse away in solution, which is correct for $t_{rebind} = 1/(N\,\max(k_1(x|6))) \sim (0.14 \text{ s})/ N$, $t_{diffuse} = \epsilon^2/D_{aq}^{EB} \sim 0.0058$ s, and $N\lesssim 20$ using the parameters from tables \ref{tab:ExperimentalModelValues} \& \ref{tab:ModifiedModelValues} and eq.(\ref{eq:SingleEBRates}).

The simulations were implemented in MATLAB using an adapted form of the Gillespie algorithm \cite{Gillespie1976, Gillespie2007} dubbed the `direct-family' method. Following cargo initialisation in the $n=1$ state at time $t_0=0$, this form of the Gillespie algorithm has been implemented as follows:

\begin{enumerate}
    \item Randomly select the time $\tau$ after which the next event occurs from the distribution $P(\tau|x_a(t_j))=k_t(x_a(t_j))\, \textnormal{exp}(-k_t(x_a(t_j))\, \tau)$ using inverse transform sampling \cite{Devroye1986}, where $k_t(x_a(t_j))$ is the total rate of any event occurring for a given cargo centre position $x_a(t_j)$ at time $t_j$;
    \item Increment the simulation time by the randomly generated time, such that $t_{j+1}=t_j + \tau$ (the index $j$ counts the total number of events that have occurred so far in the simulation);
    \item Randomly select which of the families of possible events the event that has occurred belongs to by calculating the fractional probability of each family of events occurring $P_m(x_a(t_j))=k_m(x_a(t_j))/k_t(x_a(t_j))$, where $k_m(x_a(t_j))$ is the total rate for all events from family $m$ for a given cargo centre position;
    \item Randomly select the position at which the event has occurred for the selected family, for example where the next leg binds during a binding transition by calculating the fractional probability of it being added at each position $P(x_l^+|\tau,x_a(t_j),m=1)=k_1(x_l^+)/\sum_{i=(x_a(t_j)-L)/\Delta x}^{i=(x_a(t_j)+L)/\Delta x}k_1(x_i)$ for $x_l^+ \in [x_a(t_j)-L,x_a(t_j)+L]$, or where a leg has unbound during an unbinding transition by calculating $P(x_l^-|\tau,x_a(t_j),m=2)=k_2(x_l^-)/\sum_{l=1}^{l=n} k_2(x_a(t_j))$ for $x_l^- \in \{ x_l \} (t_j)$;
    \item Update system variables based on the selected event occurring at the selected position.
\end{enumerate}

\noindent The simulations in this work require $M = 3$ different families of possible events. Importantly, each loop of this new form of the Gillespie algorithm always requires the generation of $3$ random numbers and the calculation of a single logarithm, and so is advantageous for systems with large values of both $N$ and $M$.

\subsection*{Calculating errors for simulated data}

Errors for distributions generated from simulated data were derived using a type of bootstrapping. Random samples (with replacement) were generated of the displacements and wait-times corresponding to binding, unbinding or microtubule growth events, and the error at any point of a distribution was defined as equal to the standard deviation of $25$ distributions comprised of sampled data at that point. Each sampled data set consisted of the same number of data points as that obtained in the simulation. The error calculated in this way is minimised when the number of simulated cargo becomes large, or when the motion of the cargo is highly deterministic (for example in the presence of a large microtubule growth velocity). If the plotted distribution was smoothed, smoothing was carried out before calculating the standard deviation of the samples.

\clearpage

\section*{Supplementary Figures}

\begin{figure*}[h!]
\centering
	\includegraphics[width=1.0\linewidth]{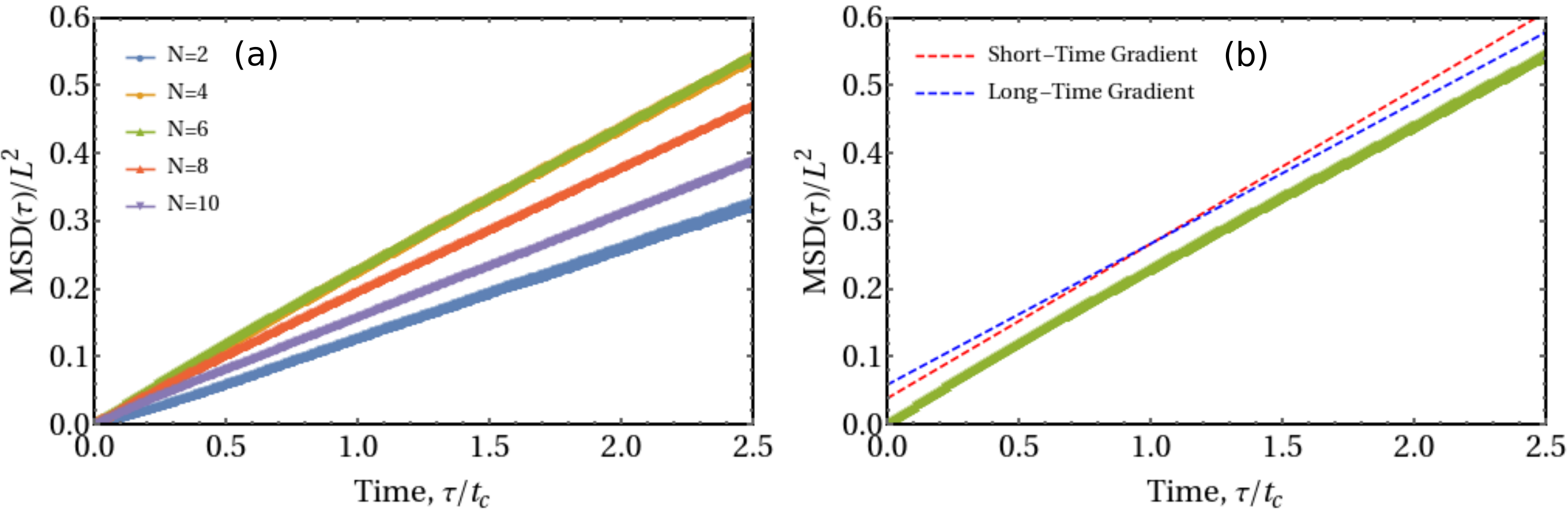}
 	\caption{(a) Cargo mean-squared displacement evolves approximately linearly with time since bound $\tau$. (b) Example mean-squared displacement distribution showing the short- and long-time gradients for $6$-legged cargo.}
 	\label{fig:MSD}
\end{figure*}

\begin{figure*}[h!]
\centering
	\includegraphics[width=0.75\linewidth]{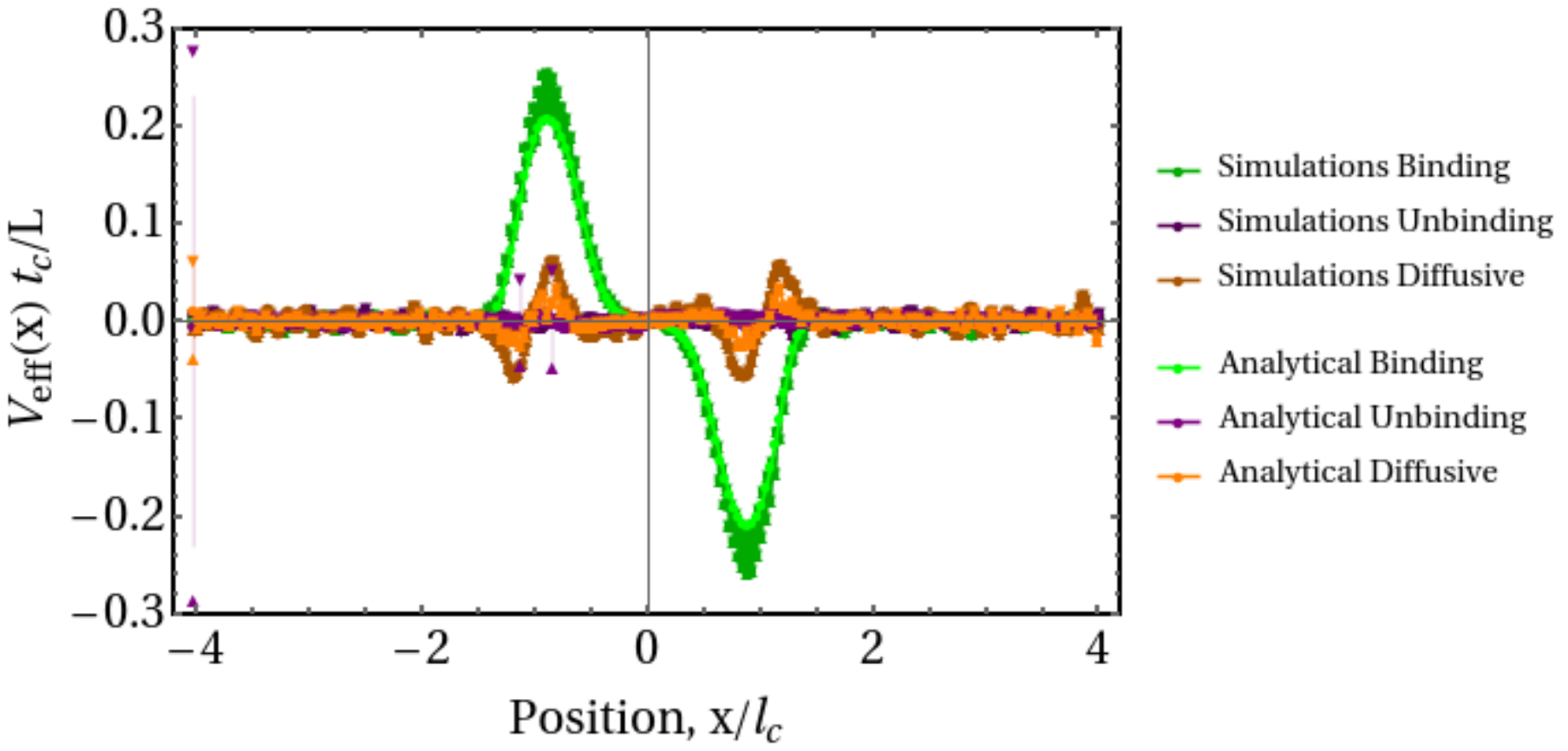}
 	\caption{Individual components of the effective velocity distributions shown in Fig.(\ref{fig:Velocity}a). The effective velocity is dominated by the component due to binding events for both the simulations and analytics.}
 	\label{fig:VelocityComponents}
\end{figure*}

\begin{figure*}[h!]
\centering
	\includegraphics[width=1.0\linewidth]{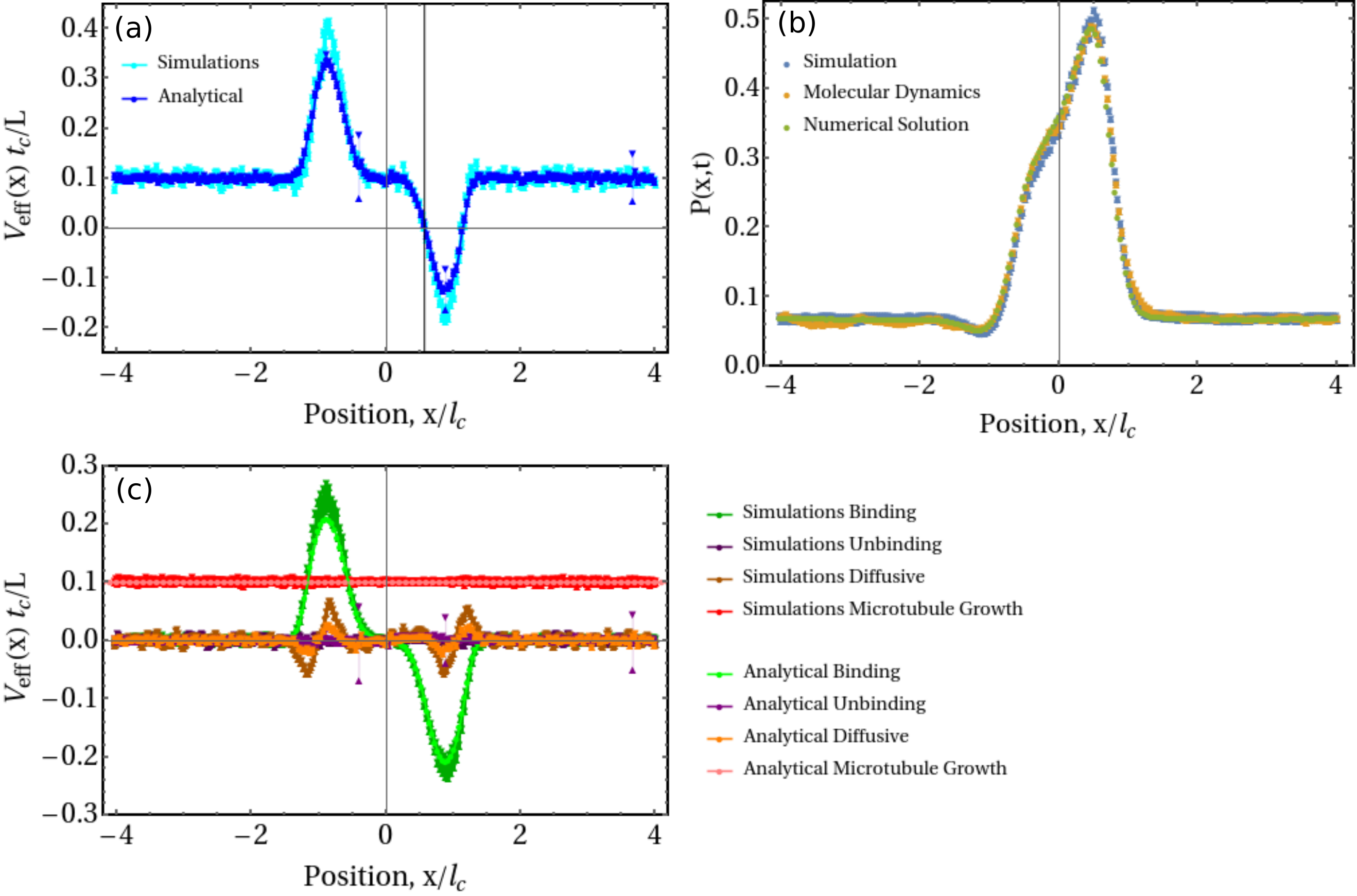}
 	\caption{(a) Introducing $\bar{k}_3\,\Delta x \neq 0$ results in cargo exhibiting a stable fixed point (black vertical line), defined as the position where $S_\lambda^{(1)}(x)=0$ and $\partial S_\lambda^{(1)}(x)/\partial x < 0$ (see eq.(\ref{eq:StableFixedPoint})). (b) Introducing $\bar{k}_3\,\Delta x \neq 0$ skews the PDF describing cargo positions towards the edge of the central region of increased binding rate defined by eq.(\ref{eq:PosDepRates}). (c) Components of the effective velocity in (a), showing the position independent microtubule growth velocity (otherwise the same as Fig.(\ref{fig:VelocityComponents})).}
 	\label{fig:NonZeroVMT}
\end{figure*}

\clearpage

\section*{Supplementary Tables}

\begin{table}[h!]
\begin{center}
\begin{tabular}{ | p{9.25cm} | p{5.75cm} | }
\hline
Parameter & Representative Literature Value \\ 
\hline
Average EB protein dwell time at microtubule ends, $\tau^{tip}$ & $(0.34 \pm 0.04)$ s \cite{Gouveia2010} \\ 
\hline
Average EB protein binding rate at microtubule ends, $k_{on}^{tip}$ & $(6.5 \pm 0.5)$ $\textnormal{nM}^{-1}\mu\textnormal{m}^{-1}\textnormal{s}^{-1}$ \cite{Gouveia2010} \\ 
\hline
Average EB protein dwell time on the microubule lattice, $\tau^{lat}$ & $(0.16 \pm 0.03)$ s \cite{Gouveia2010} \\ 
\hline
Average EB protein binding rate on the microtubule lattice, $k_{on}^{lat}$ & $(2.3 \pm 0.9)$ $\textnormal{nM}^{-1}\mu\textnormal{m}^{-1}\textnormal{s}^{-1}$ \cite{Gouveia2010} \\ 
\hline
Average formation rate of GDP-Pi tubulin sites, $k_f$ & $6.5$ s${}^{-1}$ \cite{Maurer2014} \\
\hline
Average rate of the step in the GTPase cycle converting tubulin from the GDP-Pi state to having only an associated GDP molecule, $k_h$ & $(0.23 \pm 0.01)$ s${}^{-1}$ \cite{Maurer2014} \\
\hline
The same rate for tubulin with a bound EB, $k_{EBh}$ & $(0.73 \pm 0.01)$ s${}^{-1}$ \cite{Maurer2014} \\
\hline
EB concentration, $[EB]$ & $50$ nM \cite{Maurer2014} \\ 
\hline
Average microtubule growth speed, $|v_{MT}|$ & $57$ nm$\,\textnormal{s}^{-1}$ \cite{Roostalu2020} \\
\hline
Average size of microtubule tip region, $L_{tip}$ & $220$ nm \cite{Roostalu2020} \\
\hline
Average tubulin dimer length, $\Delta x$ (distance between EB protein binding sites) & $8.185$ nm \cite{Zhang2018} \\
\hline
Microtubule radius, $\rho_{MT}$ & $12$ nm \cite{Lodish2000} \\
\hline
Bead size, $L_b$ & $\sim 15-20$ nm \cite{Rodriguez2020} \\
\hline
Radius of gyration of EB3, $\rho_{EB}$ & $4.52$ nm \cite{Buey2011} \\
\hline
Maximum diameter of EB3, $\rho_{EB}^{max}$ & $13.6$ nm \cite{Buey2011} \\
\hline
EB Diffusivity in solution, $D_{aq}^{EB}$ & $74.1$ $\mu$m${}^2$ s${}^{-1}$ \\
\hline
\end{tabular}
\end{center}

\caption{Experimental model parameters obtained from previously published work. Average dwell times are for wild-type, dimeric EB3 \cite{Gouveia2010}. Average EB binding and unbinding rates are approximately independent of the background EB concentration and the microtubule growth speed \cite{Song2020}. Microtubule growth speed is assumed to be independent of the EB concentration \cite{Maurer2014,Roostalu2020}.The quoted EB concentration was that which generates the quoted rates $k_{f,h,EBh}$ \cite{Maurer2014} and results in the microtubule growth speed closest to the presented value \cite{Maurer2014, Roostalu2020}. The size of the microtubule tip region is a function of the microtubule growth speed and hence tubulin concentration, and one set of in vitro values is quoted here \cite{Roostalu2020}. Average tubulin dimer length was obtained using cryo-electron microscopy, and the value presented is the average of the GDP lattice ($8.176$ nm) and GTP$\gamma$S lattice ($8.193$ nm) values \cite{Zhang2018}. Cargo size depends on the form of the construct, but previously published work has shown that quantum dots with the quoted radisu can exhibit EB-mediated tip tracking transport \cite{Rodriguez2020}. The radius of gyration and maximum diameter of an EB3 homodimer were obtained using small-angle X-ray scattering experiments \cite{Buey2011}. All values without errors were quoted in this form when published. The EB diffusivity was calculated using the Einstein relation assuming a temperature of $37$ ${}^o$C, that the cytoplasmic viscosity is approximately equal to that of water at $37$ ${}^o$C, and that EBs are approximately spherical with a radius equal to $\rho_{EB}$. The calculated diffusivity is of a similar order of magnitude to values measured for different biological molecules \cite{Schavemaker2018}. \label{tab:ExperimentalModelValues}}
\end{table}

\begin{table}[h!]
\begin{center}
\begin{tabular}{ | p{7.5cm} | p{3.75cm} | p{3.75cm} | }
\hline
Parameter & Minimum Value & Maximum Value \\ 
\hline
Average EB protein unbinding rate at microtubule ends, $k_{off}^{GDP-Pi}$ & $(2.9 \pm 0.3) \, \textnormal{s}^{-1}$ & $(2.9 \pm 0.3) \, \textnormal{s}^{-1}$ \\ 
\hline
Average EB protein binding rate at microtubule ends, $\kappa_{on}^{GDP-Pi}$ & $(0.0231 \pm 0.0018) \, \textnormal{s}^{-1}$ & $(27000 \pm 2000) \, \textnormal{s}^{-1}$ \\ 
\hline
Average EB protein unbinding rate from the microubule lattice, $k_{off}^{GDP}$ & $(6.3 \pm 1.0) \, \textnormal{s}^{-1}$ & $(6.3 \pm 1.0) \, \textnormal{s}^{-1}$ \\ 
\hline
Average EB protein binding rate to the microtubule lattice, $\kappa_{on}^{GDP}$ & $(0.008 \pm 0.003) \, \textnormal{s}^{-1}$ & $(9000 \pm 4000) \, \textnormal{s}^{-1}$ \\ 
\hline
\end{tabular}
\end{center}

\caption{Experimentally derived EB binding and unbinding rates from table \ref{tab:ExperimentalModelValues} converted to units of per EB per unit time using eq.(\ref{eq:Conversion}). Unbinding rates do not require conversion. The range of converted binding rates for the microtubule tip and lattice each span approximately six orders of magnitude. \label{tab:ModifiedModelValues}}
\end{table}

\clearpage

\footnotesize

\bibliographystyle{unsrt}
\bibliography{bib}{}

\begin{thebibliography}{10}

\bibitem{Perl2011}
A.~Perl, A.~Gomez-Casado, D.~Thompson, H.H. Dam, P.~Jonkheijm, D.N. Reinhoudt,
  and J.~Huskens.
\newblock Gradient-driven motion of multivalent ligand molecules along a
  surface functionalized with multiple receptors.
\newblock {\em Nat. Chem.}, 3:317--322, 2011.

\bibitem{Applewhite2010}
D.A. Applewhite, K.D. Grode, D.~Keller, A.~Zadeh, K.C. Slep, and S.L. Rogers.
\newblock The spectraplakin short stop is an actin-microtubule cross-linker
  that contributes to organization of the microtubule network.
\newblock {\em Mol. Biol. Cell}, 21(10):1714--1724, 2010.

\bibitem{Lopez2014}
M.P. L\'{o}pez, F.~Huber, I.~Grigoriev, M.O. Steinmetz, A.~Akhmanova, G.H.
  Koenderink, and M.~Dogterom.
\newblock Actin-microtubule coordination at growing microtubule ends.
\newblock {\em Nat. Commun.}, 5(4778):1--9, 2014.

\bibitem{Forth2014}
S.~Forth, K.C. Hsia, Y.~Shimamoto, and T.M. Kapoor.
\newblock Asymmetric friction of nonmotor maps can lead to their directional
  motion in active microtubule networks.
\newblock {\em Cell}, 157(2):420--432, 2014.

\bibitem{Alkemade2021}
C.~Alkemade, H.~Wierenga, V.A. Volkov, M.~Preciado-L{\/o}pez, P.R. ten~Wolde
  A.~Akhmanova, M.~Dogterom, and G.H. Koenderink.
\newblock Cross-linkers at growing microtubule ends generate forces that drive
  actin transport.
\newblock bioRxiv doi: 10.1101/2021.07.09.451744, 2021.

\bibitem{Gorbsky1987}
G.J. Gorbsky, P.J. Sammak, and G.G. Borisy.
\newblock Chromosomes move poleward in anaphase along stationary microtubules
  that coordinately disassemble from their kinetochore ends.
\newblock {\em J. Cell Biol.}, 104(1):9--18, 1987.

\bibitem{Volkov2018}
V.A. Volkov, P.J. {Huis in 't Veld}, M.~Dogterom, and A.~Musacchio.
\newblock Multivalency of ndc80 in the outer kinetochore is essential to track
  shortening microtubules and generate forces.
\newblock {\em eLife}, 7:e36764, 2018.

\bibitem{WatermanStorer1995}
C.M. Waterman-Storer, J.~Gregory, S.F. Parsons, and E.D. Salmon.
\newblock Membrane/microtubule tip attachment complexes ({TACs}) allow the
  assembly dynamics of plus ends to push and pull membranes into
  tubulovesicular networks in interphase {Xenopus} egg extracts.
\newblock {\em J. Cell Biol.}, 130(5):1161--1169, 1995.

\bibitem{WatermanStorer1998}
C.M. Waterman-Storer and E.D. Salmon.
\newblock Endoplasmic reticulum membrane tubules are distributed by
  microtubules in living cells using three distinct mechanisms.
\newblock {\em Curr. Biol.}, 8(14):798--806, 1998.

\bibitem{Grigoriev2008}
I.~Grigoriev, S.M. Gouveia, B.~van~der Vaart, J.~Demmers, J.T. Smyth,
  S.~Honnappa, D.~Splinter~M.O. Steinmetz, J.W.~Putney Jr, C.C. Hoogenraad, and
  A.~Akhmanova.
\newblock {STIM}1 is a {MT}-plus-end-tracking protein involved in remodeling of
  the {ER}.
\newblock {\em Curr. Biol.}, 18(3):177--182, 2008.

\bibitem{Nogales1999}
E.~Nogales, M.~Whittaker, R.A. Milligan, and K.H. Downing.
\newblock High-resolution model of the microtubule.
\newblock {\em Cell}, 96(1):79--88, 1999.

\bibitem{Su1995}
L.K. Su, M.~Burrell, D.E. Hill, J.~Gyuris, R.~Brent, R.~Wiltshire, J.~Trent,
  B.~Vogelstein, and K.W. Kinzler.
\newblock {APC} binds to the novel protein {EB}1.
\newblock {\em Cancer Res.}, 55(14):2972--2977, 1995.

\bibitem{Beinhauer1997}
J.D. Beinhauer, I.M. Hagan, J.H. Hegemann, and U.~Fleig.
\newblock Mal3, the fission yeast homologue of the human {APC}-interacting
  protein {EB}-1 is required for microtubule integrity and the maintenance of
  cell form.
\newblock {\em J. Cell Biol.}, 139(3):717--728, 1997.

\bibitem{Renner1997}
C.~Renner, J.P. Pfitzenmeier, K.~Gerlach, G.~Held, S.~Ohnesorge, U.~Sahin, and
  S.~Bauerand~M. Pfreundschuh.
\newblock {RP}1, a new member of the adenomatous polyposis coli-binding
  {EB}1-like gene family, is differentially expressed in activated {T} cells.
\newblock {\em J. Immunol.}, 159(3):1276--1283, 1997.

\bibitem{Tirnnauer1999}
J.S. Tirnauer, E.~O’Toole, L.~Berrueta, B.E. Bierer, and D.~Pellman.
\newblock Yeast {Bim}1p promotes the {G}1-specific dynamics of microtubules.
\newblock {\em J. Cell Biol.}, 145(5):993--1007, 1999.

\bibitem{Nakagawa2000}
H.~Nakagawa, K.~Koyama, Y.~Murata, M.~Morito, T.~Akiyama, and Y.~Nakamura.
\newblock {EB}3, a novel member of the {EB}1 family preferentially expressed in
  the central nervous system, binds to a {CNS}-specific {APC} homologue.
\newblock {\em Oncogene}, 19(2):210--216, 2000.

\bibitem{Su2001}
L.K. Su and Y.~Qi.
\newblock Characterization of human {MAPRE} genes and their proteins.
\newblock {\em Genomics}, 71(2):142--149, 2001.

\bibitem{Bieling2007}
P.~Bieling, L.~Laan, H.~Schek, E.L. Munteanu, L.~Sandblad, M.~Dogterom,
  D.~Brunner, and T.~Surrey.
\newblock Reconstitution of a microtubule plus-end tracking system in vitro.
\newblock {\em Nature}, 450(7172):1100--1105, 2007.

\bibitem{Bieling2008}
P.~Bieling, S.~Kandels-Lewis, I.A. Telley, J.~van Dijk, C.~Janke, and
  T.~Surrey.
\newblock {CLIP}-170 tracks growing microtubule ends by dynamically recognizing
  composite {EB}1/tubulin-binding sites.
\newblock {\em J. Cell Biol.}, 183(7):1223--1233, 2008.

\bibitem{Honnappa2009}
S.~Honnappa, S.M. Gouveia, A.~Weisbrich, F.F. Damberger~N.S. Bhavesh,
  H.~Jawhari, I.~Grigoriev, F.J. van Rijssel, R.M. Buey, A.~Lawera,
  I.~Jelesarov, F.K. Winkler, K.~W{\"u}thrich, A.~Akhmanova, and M.O.
  Steinmetz.
\newblock An {EB}1-binding motif acts as a microtubule tip localization signal.
\newblock {\em Cell}, 138(2):366--376, 2009.

\bibitem{Gouveia2010}
S.~Montenegro Gouveia, K.~Leslie, L.C. Kapitein, R.M. Buey, I.~Grigoriev,
  M.~Wagenbach, I.~Smal, E.~Meijering, C.C. Hoogenraad, L.~Wordeman, M.O.
  Steinmetz, and A.~Akhmanova.
\newblock In vitro reconstitution of the functional interplay between {MCAK}
  and {EB}3 at microtubule plus ends.
\newblock {\em Curr. Biol.}, 20(19):1717--1722, 2010.

\bibitem{Jiang2012}
K.~Jiang, G.~Toedt, S.~Montenegro Gouveia, N.E. Davey, S.~Hua, B.~Van~Der
  Vaart, I.~Grigoriev, J.~Larsen, L.B. Pedersen, K.~Bezstarosti,
  M.~Lince-Faria, J.~Demmers, M.O. Steinmetz, T.J. Gibson, and A.~Akhmanova.
\newblock A proteome-wide screen for mammalian {SxIP} motif-containing
  microtubule plus-end tracking proteins.
\newblock {\em Curr. Biol.}, 22(19):1800--1807, 2012.

\bibitem{Maurer2012}
P.M. Maurer, F.J. Fourniol, G.~Bohner, C.A. Moores, and T.~Surrey.
\newblock {EB}s recognize a nucleotide-dependent structural cap at growing
  microtubule ends.
\newblock {\em Cell}, 149(2):371--382, 2012.

\bibitem{Roth2018}
D.~Roth, B.P. Fitton, N.P. Chmel, N.~Wasiluk, and A.~Straube.
\newblock Spatial positioning of {EB} family proteins at microtubule tips
  involves distinct nucleotide-dependent binding properties.
\newblock {\em J. Cell Sci.}, 132(4):1--18, 2018.

\bibitem{Rodriguez2020}
R.~Rodr\'{i}guez-Garc\'{i}a, V.A. Volkov, C.Y. Chen, E.A. Katrukha, N.~Olieric,
  A.~Aher, I.~Grigoriev, M.P. L\'{o}pez, M.O. Steinmetz, L.C. Kapitein,
  G.~Koenderink, M.~Dogterom, and A.~Akhmanova.
\newblock Mechanisms of motor-independent membrane remodeling driven by dynamic
  microtubules.
\newblock {\em Curr. Biol.}, 30(6):972--987, 2020.

\bibitem{Zaytsev2013}
A.V. Zaytsev, F.I. Ataullakhanov, and E.L. Grishchuk.
\newblock Highly transient molecular interactions underlie the stability of
  kinetochore–microtubule attachment during cell division.
\newblock {\em Cel. Mol. Bioeng.}, 6:393--405, 2013.

\bibitem{Klumpp2005}
S.~Klumpp and R.~Lipowsky.
\newblock Cooperative cargo transport by several molecular motors.
\newblock {\em PNAS USA}, 102(48):17284--17289, 2005.

\bibitem{Erdmann2012}
T.~Erdmann and U.S. Schwarz.
\newblock Stochastic force generation by small ensembles of myosin {II} motors.
\newblock {\em Phys. Rev. Lett.}, 108:188101, 2012.

\bibitem{Joglekar2002}
A.P. Joglekar and A.J. Hunt.
\newblock A simple, mechanistic model for directional instability during
  mitotic chromosome movements.
\newblock {\em Biophys. J.}, 83(1):42--58, 2002.

\bibitem{Errington2019}
W.J. Errington, B.~Bruncsics, and C.A. Sarkar.
\newblock Mechanisms of noncanonical binding dynamics in multivalent
  protein–protein interactions.
\newblock {\em PNAS USA}, 116(51):25659--25667, 2019.

\bibitem{Williams2021}
I.~Tuval S.~Williams, R.~Jeanneret and M.~Polin.
\newblock Confinement-induced accumulation and spontaneous de-mixing of
  microscopic active-passive mixtures.
\newblock In preparation (2021).

\bibitem{Risken1989}
H.~Risken.
\newblock {\em The Fokker-Planck Equation, Methods of Solution and Application
  (Second Edition)}.
\newblock Springer, Berlin, 1989.

\bibitem{VanKampen1992}
N.G.~Van Kampen.
\newblock {\em Stochastic Processes in Physics and Chemistry}.
\newblock Elsevier Science B.V., Amsterdam, 1992.

\bibitem{Paul2013}
W.~Paul and J.~Baschnagel.
\newblock {\em Stochastic Processes from Physics to Finance (Second Edition)}.
\newblock Springer, Heidelberg, 2013.

\bibitem{Gillespie1976}
D.T. Gillespie.
\newblock A general method for numerically simulating the stochastic time
  evolution of coupled chemical reactions.
\newblock {\em J. Comput. Phys.}, 22(4):403--434, 1976.

\bibitem{Gillespie2007}
D.T. Gillespie.
\newblock Stochastic simulation of chemical kinetics.
\newblock {\em Annu. Rev. Phys. Chem.}, 58:35--55, 2007.

\bibitem{Lodish2000}
H.~Lodish, A.~Berk, S.L. Zipursky, P.~Matsudaira, D.~Baltimore, and J.~Darnell.
\newblock {\em Molecular cell biology}.
\newblock W.H. Freeman and Company, New York, 2000.

\bibitem{Buey2011}
R.M. Buey, R.~Mohan, K.~Leslie, T.~Walzthoeni, J.H. Missimer, A.~Menzel,
  S.~Bjeli\'{c}, K.~Bargsten, I.~Grigoriev, I.~Smal, E.~Meijering,
  R.~Aebersold, A.~Akhmanova, and M.O. Steinmetz.
\newblock Insights into {EB}1 structure and the role of its {C}-terminal domain
  for discriminating microtubule tips from the lattice.
\newblock {\em Mol. Biol. Cell}, 22(16):2912--2923, 2011.

\bibitem{Maurer2014}
P.M. Maurer, N.I. Cade, G.~Bohner, N.~Gustafsson, E.~Boutant, and T.~Surrey.
\newblock {EB}1 accelerates two conformational transitions important for
  microtubule maturation and dynamics.
\newblock {\em Curr. Biol.}, 24(4):372--384, 2014.

\bibitem{Zhang2018}
R.~Zhang, B.~Lafrance, and E.~Nogales.
\newblock Separating the effects of nucleotide and {EB} binding on microtubule
  structure.
\newblock {\em PNAS USA}, 115(27):E6191--E6200, 2018.

\bibitem{Roostalu2020}
J.~Roostalu, C.~Thomas, N.I. Cade, S.~Kunzelmann, I.A. Taylor, and T.~Surrey.
\newblock The speed of {GTP} hydrolysis determines {GTP} cap size and controls
  microtubule stability.
\newblock {\em eLife}, 9:e51992, 2020.

\bibitem{Song2020}
Y.~Song, Y.~Zhang, Y.~Pan, J.~He, Y.~Wang, W.~Chen, J.~Guo, H.~Deng, Y.~Xue,
  X.~Fang, and X.~Liang.
\newblock The microtubule end-binding affinity of {EB}1 is enhanced by a
  dimeric organization that is susceptible to phosphorylation.
\newblock {\em J. Cell Sci.}, 133(9):241216, 2020.

\bibitem{Meyhofer1995}
E.~Meyh{\"o}fer and J.~Howard.
\newblock The force generated by a single kinesin molecule against an elastic
  load.
\newblock {\em PNAS USA}, 92(2):574--578, 1995.

\bibitem{Lipka2016}
J.~Lipka, L.C Kapitein, J.~Jaworski, and C.C. Hoogenraad.
\newblock Microtubule-binding protein doublecortin-like kinase 1 ({DCLK}1)
  guides kinesin-3-mediated cargo transport to dendrites.
\newblock {\em EMBO J.}, 35(3):302--318, 2016.

\bibitem{Rickman2017}
J.~Rickman, C.~Duellberg, N.I. Cade, L.D. Griffin, and T.~Surrey.
\newblock Steady-state {EB} cap size fluctuations are determined by stochastic
  microtubule growth and maturation.
\newblock {\em J. Biol. Chem.}, 114(13):3427--3432, 2017.

\bibitem{Zwetsloot2018}
A.J. Zwetsloot, G.~Tut, and A.~Straube.
\newblock Measuring microtubule dynamics.
\newblock {\em Essays Biochem.}, 62(6):725--735, 2018.

\bibitem{Lopez2016}
B.J. Lopez and M.T. Valentine.
\newblock The +{TIP} coordinating protein {EB}1 is highly dynamic and diffusive
  on microtubules, sensitive to {GTP} analog, ionic strength, and {EB}1
  concentration.
\newblock {\em Cytoskeleton}, 73(1):23--34, 2016.

\bibitem{Devroye1986}
L.~Devroye.
\newblock {\em Non-Uniform Random Variate Generation}.
\newblock Springer, New York, 1986.

\bibitem{Schavemaker2018}
P.E. Schavemaker, A.J. Boersma, and B.~Poolman.
\newblock How important is protein diffusion in prokaryotes?
\newblock {\em Front. Mol. Biosci.}, 5(93), 2018.

\end{thebibliography}

\newpage

\end{document}